\documentclass[aps,prd,twocolumn,superscriptaddress,preprintnumbers,balancelastpage,longbibliography]{revtex4-2}

\usepackage{appendix}
\usepackage[utf8]{inputenc}
\usepackage{amsmath,amssymb,mathtools,bm}
\usepackage{graphicx, color, hepunits}
\usepackage[dvipsnames]{xcolor}
\usepackage{float}
\usepackage{multirow}
 \usepackage{hyperref} 
\hypersetup{
    colorlinks=true,       
    linkcolor=blue,        
    citecolor=blue,        
    filecolor=magenta,     
    urlcolor=blue          
}
\usepackage[utf8]{inputenc}
\usepackage[english]{babel}

\usepackage{tensor}

\begin{document}

\preprint{\hbox{CERN-TH-2023-088}}

\title{Limits from the grave: resurrecting Hitomi for decaying dark matter and \\
forecasting leading sensitivity for XRISM
}

\author{Christopher Dessert}
\affiliation{Center for Cosmology and Particle Physics, Department of Physics,
New York University, New York, NY 10003, USA}
\affiliation{Center for Computational Astrophysics, Flatiron Institute, New York, NY 10010, USA}

\author{Orion Ning}
\affiliation{Berkeley Center for Theoretical Physics, University of California, Berkeley, CA 94720, U.S.A.}
\affiliation{Theoretical Physics Group, Lawrence Berkeley National Laboratory, Berkeley, CA 94720, U.S.A.}

\author{Nicholas L. Rodd}
\affiliation{Theoretical Physics Department, CERN, 1 Esplanade des Particules, CH-1211 Geneva 23, Switzerland}

\author{Benjamin R. Safdi}
\affiliation{Berkeley Center for Theoretical Physics, University of California, Berkeley, CA 94720, U.S.A.}
\affiliation{Theoretical Physics Group, Lawrence Berkeley National Laboratory, Berkeley, CA 94720, U.S.A.}

\date{\today}

\begin{abstract}
The Hitomi X-ray satellite mission carried unique high-resolution spectrometers that were set to revolutionize the search for sterile neutrino dark matter (DM) by looking for narrow X-ray lines arising from DM decays.  Unfortunately, the satellite was lost shortly after launch, and to-date the only analysis using Hitomi for DM decay used data taken towards the Perseus cluster.  In this work we present a significantly more sensitive search from an analysis of archival Hitomi data towards blank sky locations, searching for DM decaying in our own Milky Way. The soon-to-be-launched XRISM satellite will have nearly identical soft-X-ray spectral capabilities to Hitomi; we project the full-mission sensitivity of XRISM for analyses of their future blank-sky data, and we find that XRISM will have the leading sensitivity to decaying DM for masses between roughly 1 to 20 keV, with important implications for sterile neutrino and heavy axion-like particle DM scenarios.
\end{abstract}

\maketitle

Dark matter (DM) decay is a generic prediction of many particle DM scenarios (for recent reviews, see Refs.~\cite{Boddy:2022knd,Safdi:2022xkm}). DM decays into two-body final states including a photon are especially promising discovery channels, since line-like photon signatures may stand out clearly above backgrounds across the electromagnetic spectrum. The X-ray band is a favorable energy range to look for monochromatic signatures of DM decay because of well-motivated decaying DM models in this mass range, including sterile neutrino and axion-like-particle (ALP) DM, as well as the presence of high-resolution space-based X-ray spectrometers. Moreover, the decay rates predicted by both sterile neutrino and ALP DM models are within reach of current- and next-generation instruments.

\begin{figure}[!t]
\centering
\includegraphics[width=0.5\textwidth]{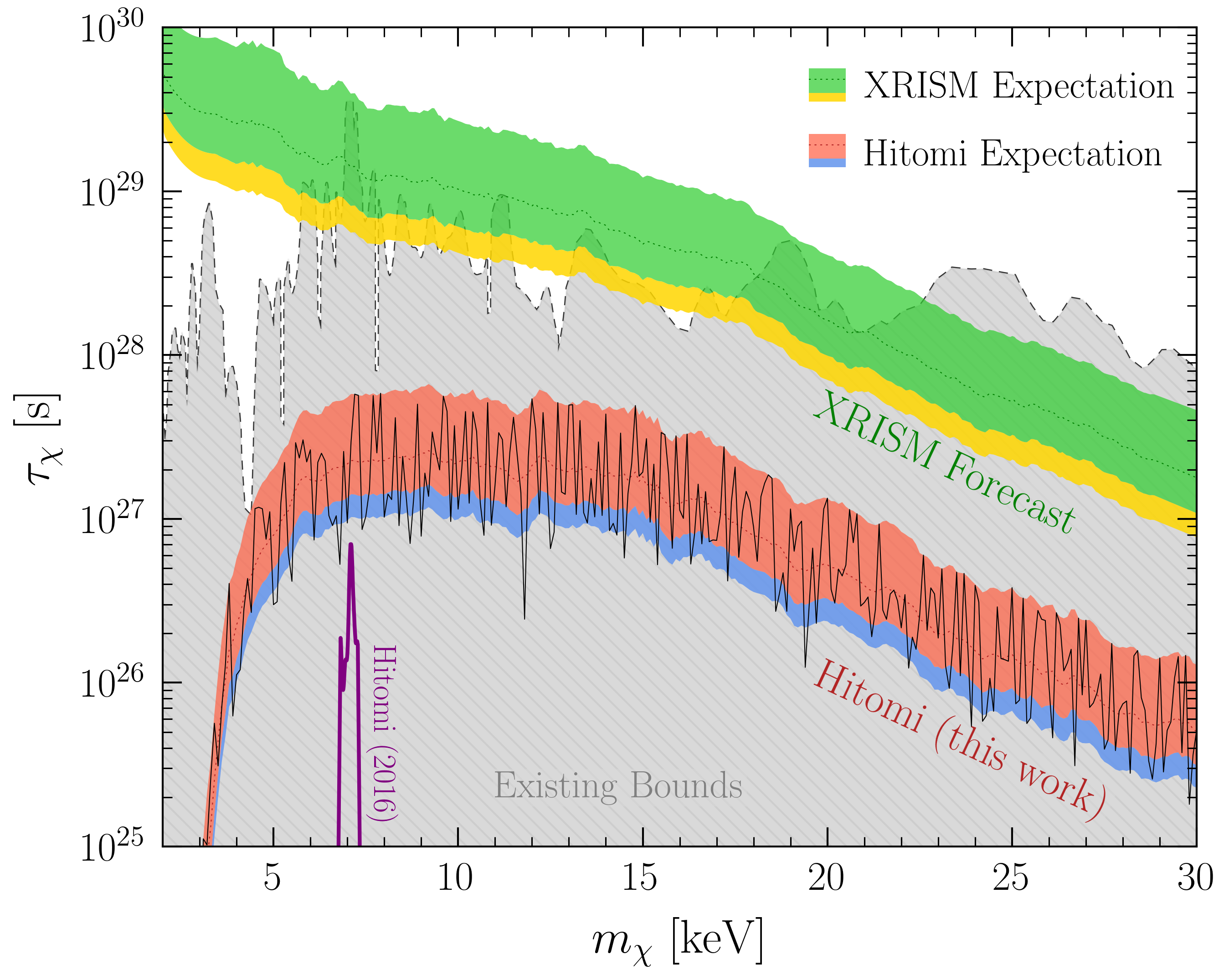}
\vspace{-0.5cm}
\caption{
The decaying DM parameter space for DM $\chi$ of mass $m_\chi$ that decays to $\chi \to \gamma + X$, where $X$ is any other final-state particle, with (partial) lifetime $\tau_\chi$. The expectation bands include the expected limit (dotted), together with the 1 and 2$\sigma$ (lower) bands. Existing limits on this parameter space are shaded in gray~\cite{Horiuchi:2013noa,Dessert:2018qih,Foster:2021ngm,Roach:2019ctw,Roach:2022lgo}, except for the Hitomi 2016 Perseus analysis upper limit that is highlighted~\cite{Hitomi:2016mun}. The Hitomi blank-sky analyses from this work substantially improve the upper limit relative to the 2016 analysis, while future analyses making use of the soon-to-be-launched XRISM satellite will set leading constraints on decaying DM over a large mass range. 
See Fig.~\ref{fig:Lifetime+Axion} for these limits recast in terms of the sterile neutrino and the ALP DM parameter spaces.}
\label{fig:Limit}
\vspace{-0.4cm}
\end{figure}

Searches for monochromatic signatures of DM decay in the X-ray band are made difficult by the fact that existing telescopes such as XMM-Newton and Chandra have energy resolutions of ${\cal O}(5\%)$, which can induce confusion between a putative DM line and astrophysical lines in the same band and which further limits the sensitivity of these instruments as the signal is smeared into the continuum backgrounds. The Hitomi instrument, on the other hand, realized an unprecedented energy resolution of ${\cal O}(0.1\%)$~\cite{2016SPIE.9905E..0VK}. Hitomi was launched on February 17, 2016, but was destroyed in orbit on March 26, 2016. Before it was lost, a small amount of data was collected, although far less than the anticipated three years of exposure. In particular, Hitomi observed the Perseus cluster; an analysis of that data in the context of decaying sterile neutrino DM in Perseus led to strong upper limits on the putative DM interaction strength with ordinary matter~\cite{Hitomi:2016hzf,Hitomi:2016mun,Tamura:2018scp}, as illustrated in Fig.~\ref{fig:Limit}.

The Perseus Hitomi analysis made use of 230\,ks of data collected by the Soft X-ray Spectrometer (SXS)~\cite{Hitomi:2016mun}.  In this work we perform an analysis of 421\,ks of Hitomi SXS blank sky data for decaying DM in the Milky Way's halo; we find no evidence for DM, and our upper limits surpass those previously derived from Perseus. While significantly improved, as Fig.~\ref{fig:Limit} demonstrates, ultimately the small Hitomi data set means our limits are subdominant compared to those derived with other instruments, though our limits may be subject to less systematic uncertainties related to background mismodeling given the improved energy resolution. The situation will soon change however. Hitomi is scheduled to be followed by the X-Ray Imaging and Spectroscopy Mission (XRISM) satellite~\cite{XRISMScienceTeam:2020rvx}, set to launch in 2023. XRISM will have nearly identical spectral capabilities to Hitomi; we show that using the full expected data set from that mission for a blank-sky search for DM will lead to leading sensitivity for decaying DM over more than a decade of possible DM masses in the 1--20 keV range. 

Currently the strongest constraints on keV-scale decaying DM come from blank-sky observations (BSOs). References~\cite{Dessert:2018qih,Dessert:2020hro,Foster:2021ngm} analyzed all archival data from the XMM-Newton telescope looking for DM decay in the ambient halo of the Milky Way to rule out the DM interpretation of the 3.5 keV line~\footnote{The 3.5 keV line is an observed unassociated X-ray line found using a variety of instruments, including XMM-Newton and Chandra, from observations of a number of targets, including the Perseus galaxy cluster, nearby galaxies such as M31, and blank regions of the Milky Way~\cite{Bulbul:2014sua,Boyarsky:2014jta,Urban:2014yda,Jeltema:2014qfa,Cappelluti:2017ywp}. Nonetheless, Refs.~\cite{Dessert:2018qih,Dessert:2020hro,Foster:2021ngm} found no evidence for the line and were able to exclude decaying DM as an explanation.}, while Refs.~\cite{Roach:2019ctw,Roach:2022lgo} used archival BSO NuSTAR data to set strong constraints on decaying DM with mass above roughly 10 keV.  The upper limits from the Hitomi analysis in this work further disfavors the DM explanation of the 3.5 keV line; while our analysis is less sensitive than previous ones around 3.5 keV, it is more accurate, given the improved spectral resolution of Hitomi relative to XMM-Newton.

The null-results on keV-mass decaying DM play a central role in the interpretation of sterile neutrino DM. These models are part of broader frameworks to explain the active neutrino masses; a sterile neutrino can generate the primordial DM abundance for $m_\chi \sim 10\,{\rm keV}$ and sterile-active mixing of order $\sin^2(2\theta) \sim 10^{-11}$, depending on resonant versus non-resonant production mechanisms and on the precise DM mass (as reviewed in Refs.~\cite{Abazajian:2017tcc,Boyarsky:2018tvu,Dasgupta:2021ies}). The mixing which generated the DM in the early Universe also allows for its decay at late times, to an (unobserved) active neutrino and a monochromatic X-ray photon with $E = m_\chi/2$~\cite{Pal:1981rm}. Because of their thermal origin, low-mass sterile neutrinos free-stream and wash out structure on small astrophysical scales; Milky Way dwarf galaxy counts claim to exclude sterile neutrinos for $m_{\chi} \gtrsim 15\,{\rm keV}$~\cite{Cherry:2017dwu,DES:2020fxi} for the conventional early Universe production mechanisms~\cite{Dodelson:1993je,Shi:1998km}, even in the presence of self-interactions amongst the active neutrinos~\cite{An:2023mkf}. Given that the active-sterile mixing angle is bounded from below in the resonant production scenario by allowing for the largest possible lepton asymmetry (see, {\it e.g.}, Refs.~\cite{DeGouvea:2019wpf,Dasgupta:2021ies}), the combination of X-ray and structure formation searches have severely narrowed the parameter space for the canonical picture of sterile neutrino DM (although see Ref.~\cite{Bringmann:2022aim}).

A scenario that is less constrained is ALPs with keV-scale masses, which have recently gained interest as motivated decaying DM candidates that can source X-ray lines (see, {\it e.g.}, Refs.~\cite{Higaki:2014zua,Jaeckel:2014qea,Foster:2022ajl,Panci:2022wlc,Langhoff:2022bij}). The ALP relic abundance may be produced either through the misalignment mechanism or through thermal scattering processes; in the misalignment case, the relic ALPs are cold regardless of the ALP mass $m_a$.  ALPs with masses $m_a < 2 m_e$, with $m_e$ the electron mass, may only decay to two photons through the interaction $\tfrac{1}{4} g_{a\gamma\gamma} a F_{\mu \nu} \tilde F^{\mu \nu}$, where $a$ is the axion and $F$ is the electromagnetic field strength tensor (with $\tilde F$ its Hodge dual). The coupling constant $g_{a\gamma\gamma}$ scales inversely with the axion decay constant $f_a$, which sets the scale for the ultraviolet completion of the theory; for $m_a$ in the keV range and $f_a$ near the grand unification scale, the axion lifetimes may be $\sim$$10^{30}$ s and within reach of current- and next-generation telescopes, such as Hitomi. Further, as shown in Ref.~\cite{Langhoff:2022bij}, strongly coupled keV ALPs make an irreducible contribution to the DM density that decays rapidly, such that it could be detected with X-ray satellites even if it only constitutes a tiny fraction of DM.

In the remainder of this Letter, we present the results of a data analysis using archival Hitomi data that produces strong constraints on decaying DM in the 1--30 keV mass range. Then, we use the Hitomi results to perform projections for end-of-mission sensitivity for the upcoming XRISM telescope, justifying the results in Fig.~\ref{fig:Limit}.

\vspace{0.2cm}
\noindent {\bf Hitomi Analysis.}
%
We reduce archival Hitomi data taken with the SXS for a total of 
9 observations towards two sources: (i) the neutron star RX J1856.5-3754, and (ii) the high-mass X-ray binary IGR J16318-4848.  (Full details of our data reduction are provided in the Supplementary Material (SM).) Both of the target point sources (PSs) produce soft X-rays, with negligible predicted X-ray emission above 1 keV when averaged over the field of view (FOV). 
We analyze data from 1.0--15.1 keV, thereby probing $m_{\chi} \in [2,\,30.2]\,{\rm keV}$, and bin the data into intervals of width 0.5\,eV. RX J1856.5-3754 (IGR J16318-4848) has an exposure of $t_{\rm exp} \simeq 171\,{\rm ks}$ ($t_{\rm exp} \simeq 250\,{\rm ks}$) and is at an angle of $17.27^\circ$ ($24.51^\circ$) from the Galactic Center (GC). 

The Hitomi SXS FOV is approximately $(2.9')^2$, corresponding to $\Delta \Omega \simeq 7 \times 10^{-7}$\,sr.  Averaged over that FOV the effective area peaks near 6\,keV input energy at a value $\sim$120\,cm$^2$.  The energy resolution steadily increases with energy, ranging from a full-width-half-max (FWHM) $\sim$4 eV at 1\,keV to $\sim$12\,eV at 15\,keV input energy.

\begin{figure}[!t]
\centering
\includegraphics[width=0.5\textwidth]{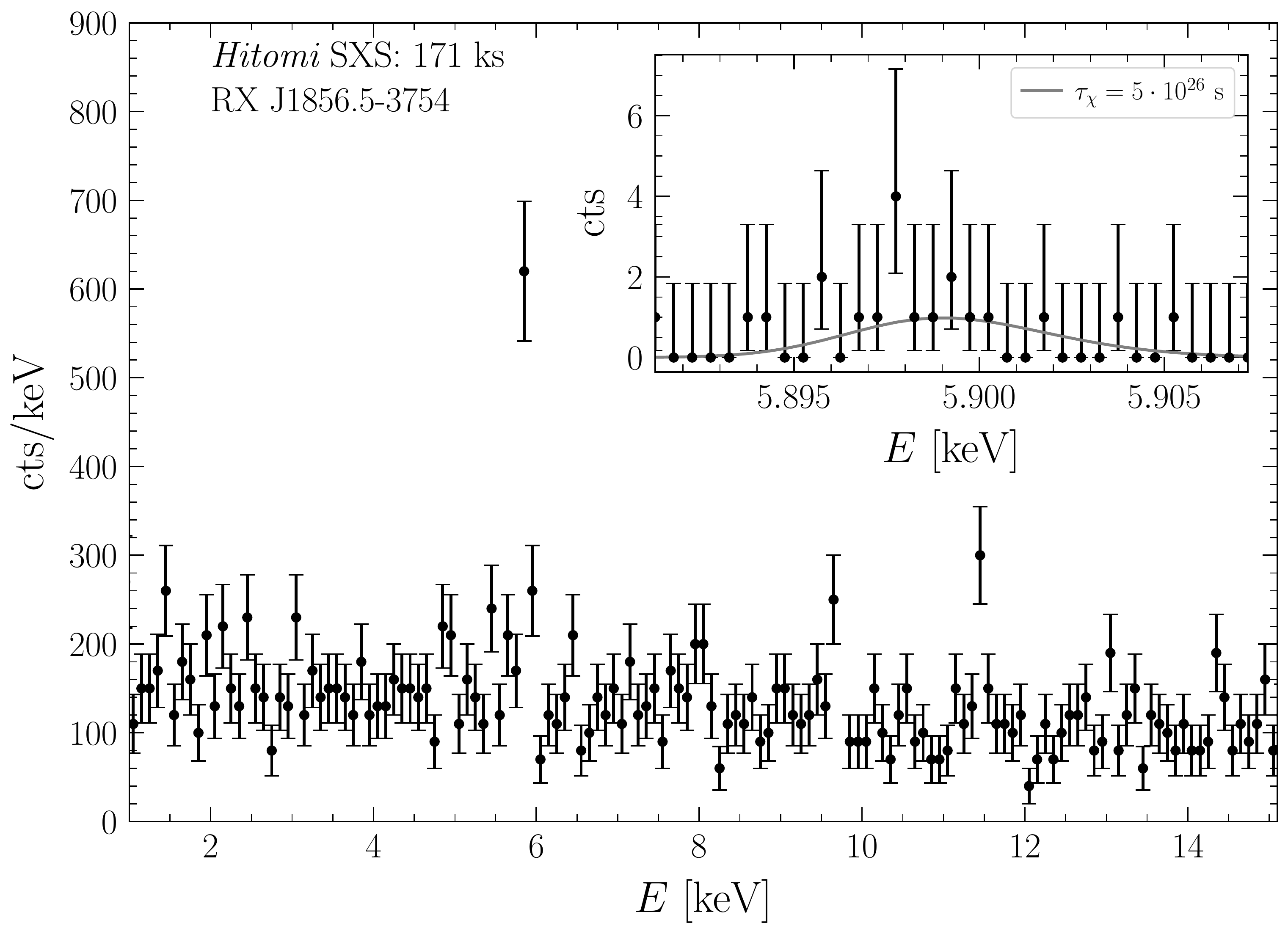}
\vspace{-0.5cm}
\caption{The stacked data for the Hitomi SXS observations towards RX J1856.5-3754, binned in 100\,eV intervals for illustration---our analysis uses 0.5\,eV bins. For the inset we focus on the most significant DM mass point for this sky location, with the data shown at the analysis-level binning and the energy range that used in the analysis. An example signal model is illustrated at the indicated lifetime.
}
\label{fig:stacked_data_1}
\vspace{-0.4cm}
\end{figure}

We stack and analyze the data separately for both pointing locations. We then combine the results of the two separate analyses using a joint likelihood, which is discussed below. In Fig.~\ref{fig:stacked_data_1} we illustrate the stacked data for the observations towards RX J1856.5-3754. For illustrative purposes we down-bin the data by a factor of 200. The data are illustrated as counts per keV with error bars 1$\sigma$ Poisson uncertainties.  In the inset of Fig.~\ref{fig:stacked_data_1} we show a zoom-in of the RX J1856.5-3754 data around the location of the highest significance excess for the DM analysis, with the data illustrated at the analysis-level energy binning of $0.5$\,eV.

For two-body DM decays within the Milky Way, the velocity dispersion of DM generates a Doppler shift that broadens the monochromatic line by $\delta E / E \sim v/c$, with $v \sim 200$\,km/s. The intrinsic width of the signal is thus expected to be $\delta E / E \sim 7 \times 10^{-4}$, which is comparable to the detector energy resolution and thus must be accounted for self consistently when searching for a decaying DM signal (see, {\it e.g.}, Ref.~\cite{Speckhard:2015eva}).  Moreover, while the DM velocity distribution is expected to be isotropic and homogeneous in the Galactic rest frame, the Sun is boosted with respect to this rest frame by ${\bf v}_\odot = {\bf v}_{\scriptscriptstyle \textrm{LSR}} + {\bf v}_{\odot, {\rm pec}}$, where ${\bf v}_{\scriptscriptstyle \textrm{LSR}} \simeq (0,220,0)\,{\rm km/s}$ tracks the local rotation velocity, and ${\bf v}_{\odot, {\rm pec}} \simeq (11,12,7)$\,km/s is the peculiar velocity of the Sun~\cite{Schoenrich:2009bx,2000A&A...354..522M}. (We work in Galactic coordinates, with ${\bf \hat x}$ pointing towards the GC, ${\bf \hat y}$ pointing in the direction of the local disk rotation, and ${\bf \hat z}$ pointing towards the north Galactic pole.) Due to our motion, pointings at $(\ell,b) = (90^\circ, 0^\circ)$ should look for higher-frequency signals than pointings at $(\ell,b) = (-90^\circ, 0^\circ)$ by $\delta E / E \sim 2 \,|{\bf v}_{\scriptscriptstyle \textrm{LSR}}| / c \sim 1.5 \times 10^{-3}$.  
To incorporate this effect, we compute the probability distribution function $f(E; m_\chi, \ell,b)$, which tells us the expected distribution of X-ray energies $E$. Full details of our procedure and illustrations of the effect are given in the SM.

From here, the differential flux from DM decay incident on the detector is
\begin{equation}
\Phi(E,\ell,b) = {1 \over 4 \pi m_\chi \tau_\chi} f(E; m_\chi, \ell,b)\, {\cal D}(\ell,b),
\end{equation}
where $\Phi$ has units of [cts/keV/cm$^2$/s/sr], $\tau_\chi^{-1} = \Gamma(\chi \to \gamma + X)$, and the astrophysical ${\cal D}$-factor is determined from ${\cal D}(\ell,b) = \int ds \rho_{\scriptscriptstyle \textrm{DM}}(r)$, with $s$ the line of sight distance and $\rho_{\scriptscriptstyle \textrm{DM}}(r)$ the DM density at a distance $r$ from the GC. Following Ref.~\cite{Foster:2021ngm}, we model $\rho_{\scriptscriptstyle \textrm{DM}}$ by a Navarro-Frenk-White (NFW) profile~\cite{Navarro:1995iw,Navarro:1996gj} with mass and scale radius parameters taken to be the most conservative values within the 68\% uncertainty range from the analysis in Ref.~\cite{Cautun:2019eaf} that constrained the DM density profile using Milky Way rotation curve data and satellite kinematic data (see the SM for specific values). The ${\cal D}$-factor at the location of RX J1856.5-3754 (IGR J16318-4848) is then calculated to be ${\cal D} \simeq 4.7 \times 10^{28}$ keV/cm$^2$ ($\simeq 3.7 \times 10^{28}$ keV/cm$^2$).

To determine the predicted signal counts Hitomi would observe, we forward model the DM flux through the instrument response as follows,
\begin{equation}
N_i^{\rm sig} = t_{\rm exp} \Delta \Omega \sum_j A^{\rm eff}_{ij}  \int_{E_j}^{E_{j+1}} dE\, \Phi(E_j,\ell,b),
\end{equation}
where $A^{\rm eff}_{ij}$ is the detector response matrix that relates input energy $E_j$ to the appropriate output $E_i$. Here $i$ and $j$ label the output and input bins, respectively, and the same energy binning is used for both. $A^{\rm eff}_{ij}$ has units of [cm$^2$].
Given a putative DM mass $m_\chi$, we model the data as a linear combination of the above signal model and a flat background model: $N_i^{\rm back} = A_{\rm back}$, treating $A_{\rm back}$ as a nuisance parameter. For fixed $m_\chi$ the DM lifetime $\tau_\chi$ is taken to be the signal model parameter of interest; note that $\tau_\chi$ is allowed to be negative, although this is unphysical, to ensure that we reach the point of maximum likelihood (see, {\it e.g.}, Refs.~\cite{Cowan:2010js,Safdi:2022xkm}). 

For a given location on the sky, we analyze all stacked data at that location using a Poisson likelihood:
\begin{equation}
p({\bm d}| {\cal M}, \{\tau_\chi, A_{\rm back}\}) = \prod_i \frac{\lambda_i^{d_i} e^{-\lambda_i}}{d_i!},
\end{equation}
where $\lambda_i \equiv N_i^{\rm back} + N_i^{\rm sig}$ is the model prediction in energy bin $i$ for the model ${\cal M}$ with parameters $\tau_\chi$ and $A_{\rm back}$, ${\bm d} = \{ d_i \}$ is the data set consisting of all stacked counts $d_i$ within the energy range of interest. In our fiducial analysis we use a sliding energy window that is centered around the peak-signal energy ($\simeq m_\chi / 2$) and includes energies within $\pm 3\sigma_E$, with $\sigma_E = {\rm FWHM}/2 \sqrt{2 \ln 2}$. We then construct the frequentist profile likelihood by maximizing the likelihood at fixed $\tau_\chi$ over $A_{\rm back}$; the joint likelihood between both locations on the sky is then given by the product of two profile likelihoods.  In the inset of Fig.~\ref{fig:stacked_data_1} we illustrate an example signal model at the indicated lifetime for the mass point with the highest significance excess; the energy range shown is that used in the analysis at that mass point.

The number of counts within the sliding analysis window summed over all observations is typically around 10, making the application of Wilks' theorem and the use of asymptotic theorems for the distribution of the discovery and upper-limit test statistics (TSs) marginally justified, so long as we restrict to TS differences less than $\sim$10 from the point of maximum likelihood~\cite{Cowan:2010js}. Note that the discovery TS is zero for negative best-fit signal strengths and is otherwise twice the difference in the log profile likelihood between the null point $\tau_\chi = 0 $ and the best-fit point $\hat \tau_\chi$; the TS for upper limits is defined similarly. 

Our largest discovery TS is $\simeq$$16$ and naively outside of the range of validity of where Wilks' theorem should hold; however, that excess appears in a region of larger-than-typical counts, and as we show explicitly in the SM through Monte Carlo (MC) simulations of the null hypothesis, the discovery TS distribution is adequately described by the one-sided chi-square distribution to the necessary precision.  We thus assume Wilks' theorem throughout this work in calculating one-sided upper limits and discovery significances. 
We test the signal hypothesis over a range of 14,100 DM mass points spanning from 2.0 keV to 30.2 keV in 2 eV intervals in order to over-resolve the detector energy resolution.  The resulting 95\% one-sided upper limit is illustrated in Fig.~\ref{fig:Limit}, along with the expected 1$\sigma$ and 2$\sigma$ containment intervals for the 95\% one-sided power-constrained upper limit under the null hypothesis~\cite{Cowan:2010js,Cowan:2011an}. (Note that power-constrained limits are not allowed to fluctuate beyond the lower 1$\sigma$ expectation for the limit under the null hypothesis.) Purely for presentation, the results in Fig.~\ref{fig:Limit} are smoothed over a mass range $0.4\,{\rm keV}$, although the unsmoothed limit is available in Ref.~\cite{supp_data}.
The limits are presented in terms of the sterile neutrino and ALP DM parameter spaces in SM Fig.~\ref{fig:Lifetime+Axion}.

\begin{figure}[!t]
\centering
\includegraphics[width=0.5\textwidth]{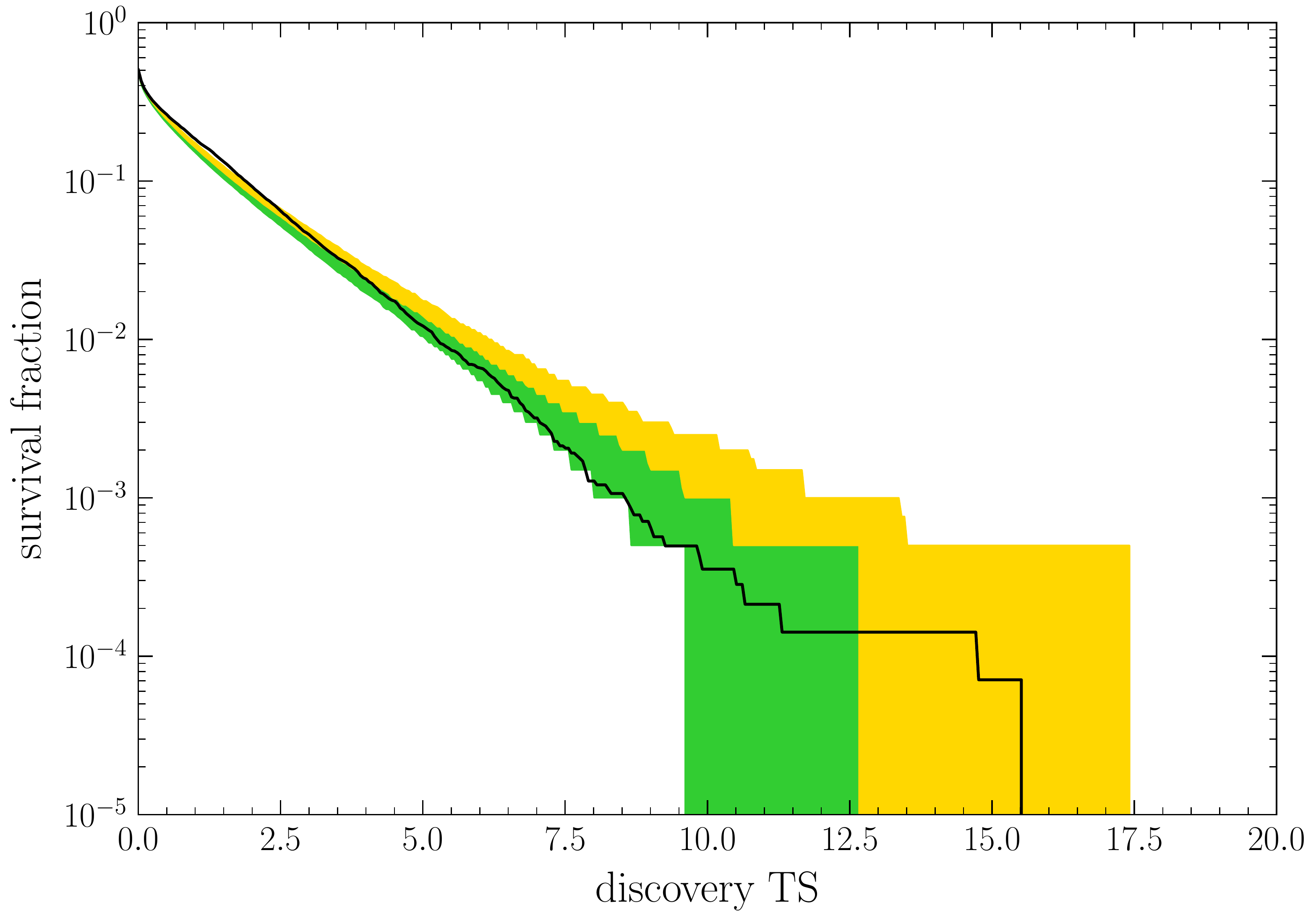}
\vspace{-0.5cm}
\caption{The survival fraction for the Hitomi analysis in this work showing the fraction of test mass points with a discovery TS at or above the value indicated on the $x$-axis.  The expectations under the null hypothesis assuming chi-square distributed TSs at 1$\sigma$ and 2$\sigma$ (upper percentile only) containment are also shown. The observed distribution of the TSs is consistent with the expectation under the null hypothesis at 68\% confidence, indicating no evidence for decaying DM.  The slight excess of low-TS points is likely due to deviations from the chi-square distribution due to low counting statistics at some test masses. }
\label{fig:survival}
\vspace{-0.4cm}
\end{figure}

In Fig.~\ref{fig:survival} we show that no high-significance excesses are observed. In detail, we show the survival fraction of discovery TSs in the data over the ensemble of all test mass points. 
That is, the figure illustrates the fraction of discovery TSs on the $y$-axis that have a TS at or above the value on the $x$-axis. The 1$\sigma$ and (upper) 2$\sigma$ expectations for the survival fraction under the null hypothesis
are illustrated in green and gold, respectively. The highest discovery TS point has a value $\sim$15.5 at $m_\chi = 11.794$ keV, which is expected within 95\% confidence under the null hypothesis over the ensemble of all mass points tested (despite corresponding to approximately $4$$\sigma$ local significance).  However, this particular high-TS test point likely corresponds to the Mn K$\alpha$ instrumental line at $5898.8010(84)$ eV~\cite{RevModPhys.75.35}; we also find ${\rm TS} \sim 4$ excesses around 6.4 keV, which could be the Fe K$\alpha_{1,2}$ lines~\cite{Hitomi:2017qip}. Thus, we conclude that the data show no evidence for decaying DM.

\vspace{0.2cm}
\noindent {\bf XRISM Projections.}
%
The {\it Resolve} instrument onboard the upcoming XRISM satellite mission is designed to have the same performance capabilities as the SXS of Hitomi~\cite{XRISMScienceTeam:2020rvx}. Thus, in making projections for XRISM we use the observed background rates from Hitomi along with the forward modeling matrices from the SXS, except that the gate valve (GV) is open. In the early calibration phase, Hitomi operated with the GV closed that severely limited the X-ray transmission at energies below a few keV. XRISM will open the GV before beginning science operations. In short, our analysis pipeline for XRISM projections assumes a detector identical to SXS but with a full mission's worth of observing time. XRISM is designed to have a three-year cryogen lifetime, though the mechanical cooling system should allow it to surpass this design goal by several years. We assume a decade-long operation, corresponding to $9.25$ years of science data, accounting for an initial nine-month calibration period. During this live-time, we assume the minimum design requirement of 90\% observing efficiency (the goal is $>$98\%). Given that we do not know where XRISM will observe, we assume that it will follow the same observing pattern as XMM-Newton. While in reality XRISM will almost certainly not follow this precise observing pattern, by basing the observations off of those from XMM-Newton we account for the slight preference to observe near the Galactic plane and near the GC in particular. The full XMM-Newton exposure distribution across the sky, as computed in Ref.~\cite{Foster:2021ngm}, is shown in the SM.

Within the eventual XRISM data set, there will be observations towards sources that have X-ray fluxes that are too bright to be useful above 1\,keV for BSOs. We use the XMM-Newton source catalog~\cite{Webb:2020rgy} to estimate that 20\% of sources have a flux above 1 keV that is more than twice the cosmic X-ray background; we assume that these sources are not included in our analysis.  In total, we thus include \mbox{$9.25 \cdot 0.9 \cdot 0.8 = 6.66$}\,yrs of data in our projections.

We follow Ref.~\cite{Foster:2021ngm} and bin the data into 30 concentric annuli centered around the GC of radial width 6$^\circ$, masking the Galactic plane for latitudes $|b| \leq 2^\circ$. In binning the data we shift the energies of the photons between different pointings to a common rest frame, accounting for the different signal offsets in energy by the Doppler shift, depending on the sky location.  We compute the profile likelihood for $\tau_\chi$ in each annulus independently for each mass point $m_\chi$, with each annulus having its own nuisance parameter $A_{\rm back}$ describing the normalization of the flat background in the sliding energy window.  We then construct the joint profile likelihood for $\tau_\chi$ as the product of the 30 profile likelihoods from the individual annuli.  
The resulting projected upper limit under the null hypothesis is illustrated in Fig.~\ref{fig:Limit}.

\vspace{0.2cm}
\noindent {\bf Discussion.}
%
With the imminent launch of XRISM it is critical to identify the optimal observing strategy for DM decay models in the X-ray band resulting in monochromatic final states. Here, we show that using BSOs to search for DM decay in the ambient halo of the Milky Way will result in significantly improved sensitivity relative to current constraints over the $\sim$1--20 keV mass range. The blank-sky analyses have the advantage of being symbiotic to existing XRISM science goals, since they do not require new, dedicated observations beyond those already planned for other reasons.

On the other hand, one could imagine performing dedicated observations towards motivated targets.  Consider, for example, observations towards the Perseus cluster and towards an especially promising dwarf galaxy~\cite{Ando:2021fhj}, such as Segue I. As illustrated in SM Fig.~\ref{fig:D}, the ${\cal D}$-factor from Perseus is roughly twice as large as that of the Milky Way's halo in the $\sim$18$^\circ$ averaged over the XRISM FOV, accounting for uncertainties. However, the X-ray background from Perseus is over roughly 100 times larger than the instrumental background, meaning that 
BSO analyses in the inner $\sim$18$^\circ$ will be at least 5 times more constraining for the same observation time relative to Perseus analyses, while also subject to less systematic uncertainties from background mismodeling.   
The Segue I ${\cal D}$-factor may be comparable to that of the Milky Way in the inner $18^\circ$, though it could also be much smaller accounting for its uncertainties.  
Additionally, we expect XRISM to have over 2 Ms of exposure within 18$^\circ$ with the first three years of science data. 
With planned observing strategies Milky Way BSO searches should provide superior sensitivity to decaying DM relative to cluster and dwarf galaxy searches.

Sterile neutrinos and keV-scale ALPs remain promising DM candidates that can lead to observable, monochromatic signatures in the X-ray band. The upcoming XRISM mission may therefore be the first to detect evidence for these models by improving the sensitivity to decaying DM across a broad range of DM masses using BSOs (see also Refs.~\cite{Lovell:2018xng,Lovell:2019aol,Lovell:2023olv}). That XRISM can be more sensitive than large scale X-ray observatories such as XMM-Newton and Chandra, which have each collected more than twenty years of data, highlights the power of the improved energy resolution for DM line searches.

In the event that XRISM does not detect a signal, it will be important to plan future missions that are better optimized for blank-sky searches. In particular, a more sensitive mission for the signal discussed in this work would have comparable energy resolution and effective area to XRISM but a much larger FOV, even if this comes at the expense of angular resolution. Indeed, as XRISM can only look at about 1 part in $10^7$ of the sky at once, while the DM signal is nearly $4\pi$ in solid angle, significant improvement should be possible in the future with wider FOV instruments that have comparable energy resolution to XRISM. The future Athena mission will provide a step in that direction with comparable energy resolution to XRISM but a modestly larger effective area and FOV~\cite{Barret:2013mxa}; in contrast, the soon-to-be-released eROSITA data set provides a complementary approach, given that it has a much larger FOV than XRISM but an energy resolution more comparable to XMM-Newton (see in particular Ref.~\cite{Dekker:2021bos}). However, with the full-data-set XMM-Newton analysis in Ref.~\cite{Foster:2021ngm} already having sizeable systematic uncertainties, systematics may dominate instruments with XMM-Newton-level energy resolution that push to deeper sensitivity by collecting more statistics.  High spectral resolution instruments such as XRISM are necessary to establish robust evidence for signals of sterile neutrinos, axions, and the DM of our Universe in the X-ray band, and blank-sky searches provide the optimal strategy to achieve this goal.

\vspace{0.5cm}
\noindent {\it Acknowledgements.}
%
We thank Aurora Simionescu for discussions on Hitomi data reduction, Joshua Foster and Yujin Park for helpful conversations, and John Beacom and Stefano Profumo for comments on the manuscript. B.R.S is supported in part by the DOE Early Career Grant DESC0019225.  The work of O.N. is supported in part by the NSF Graduate Research Fellowship Program under Grant DGE2146752.  C.D. is supported by the National Science Foundation under Grants No. 1915409 and 2210551.

\bibliography{refs}

\begin{thebibliography}{60}%
\makeatletter
\providecommand \@ifxundefined [1]{%
 \@ifx{#1\undefined}
}%
\providecommand \@ifnum [1]{%
 \ifnum #1\expandafter \@firstoftwo
 \else \expandafter \@secondoftwo
 \fi
}%
\providecommand \@ifx [1]{%
 \ifx #1\expandafter \@firstoftwo
 \else \expandafter \@secondoftwo
 \fi
}%
\providecommand \natexlab [1]{#1}%
\providecommand \enquote  [1]{``#1''}%
\providecommand \bibnamefont  [1]{#1}%
\providecommand \bibfnamefont [1]{#1}%
\providecommand \citenamefont [1]{#1}%
\providecommand \href@noop [0]{\@secondoftwo}%
\providecommand \href [0]{\begingroup \@sanitize@url \@href}%
\providecommand \@href[1]{\@@startlink{#1}\@@href}%
\providecommand \@@href[1]{\endgroup#1\@@endlink}%
\providecommand \@sanitize@url [0]{\catcode `\\12\catcode `\$12\catcode
  `\&12\catcode `\#12\catcode `\^12\catcode `\_12\catcode `\%12\relax}%
\providecommand \@@startlink[1]{}%
\providecommand \@@endlink[0]{}%
\providecommand \url  [0]{\begingroup\@sanitize@url \@url }%
\providecommand \@url [1]{\endgroup\@href {#1}{\urlprefix }}%
\providecommand \urlprefix  [0]{URL }%
\providecommand \Eprint [0]{\href }%
\providecommand \doibase [0]{https://doi.org/}%
\providecommand \selectlanguage [0]{\@gobble}%
\providecommand \bibinfo  [0]{\@secondoftwo}%
\providecommand \bibfield  [0]{\@secondoftwo}%
\providecommand \translation [1]{[#1]}%
\providecommand \BibitemOpen [0]{}%
\providecommand \bibitemStop [0]{}%
\providecommand \bibitemNoStop [0]{.\EOS\space}%
\providecommand \EOS [0]{\spacefactor3000\relax}%
\providecommand \BibitemShut  [1]{\csname bibitem#1\endcsname}%
\let\auto@bib@innerbib\@empty
\bibitem [{\citenamefont {Boddy}\ \emph {et~al.}(2022)\citenamefont {Boddy}
  \emph {et~al.}}]{Boddy:2022knd}%
  \BibitemOpen
  \bibfield  {author} {\bibinfo {author} {\bibfnamefont {K.~K.}\ \bibnamefont
  {Boddy}} \emph {et~al.},\ }\bibfield  {title} {\bibinfo {title}
  {{Snowmass2021 theory frontier white paper: Astrophysical and cosmological
  probes of dark matter}},\ }\href
  {https://doi.org/10.1016/j.jheap.2022.06.005} {\bibfield  {journal} {\bibinfo
   {journal} {JHEAp}\ }\textbf {\bibinfo {volume} {35}},\ \bibinfo {pages}
  {112} (\bibinfo {year} {2022})},\ \Eprint {https://arxiv.org/abs/2203.06380}
  {arXiv:2203.06380 [hep-ph]} \BibitemShut {NoStop}%
\bibitem [{\citenamefont {Safdi}(2022)}]{Safdi:2022xkm}%
  \BibitemOpen
  \bibfield  {author} {\bibinfo {author} {\bibfnamefont {B.~R.}\ \bibnamefont
  {Safdi}},\ }\bibfield  {title} {\bibinfo {title} {{TASI Lectures on the
  Particle Physics and Astrophysics of Dark Matter}},\ }\href@noop {} {\
  (\bibinfo {year} {2022})},\ \Eprint {https://arxiv.org/abs/2303.02169}
  {arXiv:2303.02169 [hep-ph]} \BibitemShut {NoStop}%
\bibitem [{\citenamefont {Kelley}\ \emph {et~al.}(2016)\citenamefont {Kelley}
  \emph {et~al.}}]{2016SPIE.9905E..0VK}%
  \BibitemOpen
  \bibfield  {author} {\bibinfo {author} {\bibfnamefont {R.~L.}\ \bibnamefont
  {Kelley}} \emph {et~al.},\ }\bibfield  {title} {\bibinfo {title} {{The
  Astro-H high resolution soft x-ray spectrometer}},\ }in\ \href
  {https://doi.org/10.1117/12.2232509} {\emph {\bibinfo {booktitle} {Space
  Telescopes and Instrumentation 2016: Ultraviolet to Gamma Ray}}},\ \bibinfo
  {series} {Society of Photo-Optical Instrumentation Engineers (SPIE)
  Conference Series}, Vol.\ \bibinfo {volume} {9905},\ \bibinfo {editor}
  {edited by\ \bibinfo {editor} {\bibfnamefont {J.-W.~A.}\ \bibnamefont {{den
  Herder}}}, \bibinfo {editor} {\bibfnamefont {T.}~\bibnamefont
  {{Takahashi}}},\ and\ \bibinfo {editor} {\bibfnamefont {M.}~\bibnamefont
  {{Bautz}}}}\ (\bibinfo {year} {2016})\ p.\ \bibinfo {pages}
  {99050V}\BibitemShut {NoStop}%
\bibitem [{\citenamefont {Aharonian}\ \emph {et~al.}(2016)\citenamefont
  {Aharonian} \emph {et~al.}}]{Hitomi:2016hzf}%
  \BibitemOpen
  \bibfield  {author} {\bibinfo {author} {\bibfnamefont {F.}~\bibnamefont
  {Aharonian}} \emph {et~al.} (\bibinfo {collaboration} {Hitomi}),\ }\bibfield
  {title} {\bibinfo {title} {{The Quiescent Intracluster Medium in the Core of
  the Perseus Cluster}},\ }\href {https://doi.org/10.1038/nature18627}
  {\bibfield  {journal} {\bibinfo  {journal} {Nature}\ }\textbf {\bibinfo
  {volume} {535}},\ \bibinfo {pages} {117} (\bibinfo {year} {2016})},\ \Eprint
  {https://arxiv.org/abs/1607.04487} {arXiv:1607.04487 [astro-ph.GA]}
  \BibitemShut {NoStop}%
\bibitem [{\citenamefont {Aharonian}\ \emph {et~al.}(2017)\citenamefont
  {Aharonian} \emph {et~al.}}]{Hitomi:2016mun}%
  \BibitemOpen
  \bibfield  {author} {\bibinfo {author} {\bibfnamefont {F.~A.}\ \bibnamefont
  {Aharonian}} \emph {et~al.} (\bibinfo {collaboration} {Hitomi}),\ }\bibfield
  {title} {\bibinfo {title} {{$Hitomi$ constraints on the 3.5 keV line in the
  Perseus galaxy cluster}},\ }\href {https://doi.org/10.3847/2041-8213/aa61fa}
  {\bibfield  {journal} {\bibinfo  {journal} {Astrophys. J. Lett.}\ }\textbf
  {\bibinfo {volume} {837}},\ \bibinfo {pages} {L15} (\bibinfo {year}
  {2017})},\ \Eprint {https://arxiv.org/abs/1607.07420} {arXiv:1607.07420
  [astro-ph.HE]} \BibitemShut {NoStop}%
\bibitem [{\citenamefont {Tamura}\ \emph {et~al.}(2019)\citenamefont {Tamura}
  \emph {et~al.}}]{Tamura:2018scp}%
  \BibitemOpen
  \bibfield  {author} {\bibinfo {author} {\bibfnamefont {T.}~\bibnamefont
  {Tamura}} \emph {et~al.},\ }\bibfield  {title} {\bibinfo {title} {{An X-ray
  spectroscopic search for dark matter and unidentified line signatures in the
  Perseus cluster with Hitomi}},\ }\href {https://doi.org/10.1093/pasj/psz023}
  {\bibfield  {journal} {\bibinfo  {journal} {Publ. Astron. Soc. Jap.}\
  }\textbf {\bibinfo {volume} {71}},\ \bibinfo {pages} {Publications of the
  Astronomical Society of Japan, Volume 71, Issue 3, June 2019, 50,
  https://doi.org/10.1093/pasj/psz023} (\bibinfo {year} {2019})},\ \Eprint
  {https://arxiv.org/abs/1811.05767} {arXiv:1811.05767 [astro-ph.HE]}
  \BibitemShut {NoStop}%
\bibitem [{\citenamefont {Horiuchi}\ \emph {et~al.}(2014)\citenamefont
  {Horiuchi}, \citenamefont {Humphrey}, \citenamefont {Onorbe}, \citenamefont
  {Abazajian}, \citenamefont {Kaplinghat},\ and\ \citenamefont
  {Garrison-Kimmel}}]{Horiuchi:2013noa}%
  \BibitemOpen
  \bibfield  {author} {\bibinfo {author} {\bibfnamefont {S.}~\bibnamefont
  {Horiuchi}}, \bibinfo {author} {\bibfnamefont {P.~J.}\ \bibnamefont
  {Humphrey}}, \bibinfo {author} {\bibfnamefont {J.}~\bibnamefont {Onorbe}},
  \bibinfo {author} {\bibfnamefont {K.~N.}\ \bibnamefont {Abazajian}}, \bibinfo
  {author} {\bibfnamefont {M.}~\bibnamefont {Kaplinghat}},\ and\ \bibinfo
  {author} {\bibfnamefont {S.}~\bibnamefont {Garrison-Kimmel}},\ }\bibfield
  {title} {\bibinfo {title} {{Sterile neutrino dark matter bounds from galaxies
  of the Local Group}},\ }\href {https://doi.org/10.1103/PhysRevD.89.025017}
  {\bibfield  {journal} {\bibinfo  {journal} {Phys. Rev.}\ }\textbf {\bibinfo
  {volume} {D89}},\ \bibinfo {pages} {025017} (\bibinfo {year} {2014})},\
  \Eprint {https://arxiv.org/abs/1311.0282} {arXiv:1311.0282 [astro-ph.CO]}
  \BibitemShut {NoStop}%
\bibitem [{\citenamefont {Dessert}\ \emph
  {et~al.}(2020{\natexlab{a}})\citenamefont {Dessert}, \citenamefont {Rodd},\
  and\ \citenamefont {Safdi}}]{Dessert:2018qih}%
  \BibitemOpen
  \bibfield  {author} {\bibinfo {author} {\bibfnamefont {C.}~\bibnamefont
  {Dessert}}, \bibinfo {author} {\bibfnamefont {N.~L.}\ \bibnamefont {Rodd}},\
  and\ \bibinfo {author} {\bibfnamefont {B.~R.}\ \bibnamefont {Safdi}},\
  }\bibfield  {title} {\bibinfo {title} {{The dark matter interpretation of the
  3.5-keV line is inconsistent with blank-sky observations}},\ }\href
  {https://doi.org/10.1126/science.aaw3772} {\bibfield  {journal} {\bibinfo
  {journal} {Science}\ }\textbf {\bibinfo {volume} {367}},\ \bibinfo {pages}
  {1465} (\bibinfo {year} {2020}{\natexlab{a}})},\ \Eprint
  {https://arxiv.org/abs/1812.06976} {arXiv:1812.06976 [astro-ph.CO]}
  \BibitemShut {NoStop}%
\bibitem [{\citenamefont {Foster}\ \emph {et~al.}(2021)\citenamefont {Foster},
  \citenamefont {Kongsore}, \citenamefont {Dessert}, \citenamefont {Park},
  \citenamefont {Rodd}, \citenamefont {Cranmer},\ and\ \citenamefont
  {Safdi}}]{Foster:2021ngm}%
  \BibitemOpen
  \bibfield  {author} {\bibinfo {author} {\bibfnamefont {J.~W.}\ \bibnamefont
  {Foster}}, \bibinfo {author} {\bibfnamefont {M.}~\bibnamefont {Kongsore}},
  \bibinfo {author} {\bibfnamefont {C.}~\bibnamefont {Dessert}}, \bibinfo
  {author} {\bibfnamefont {Y.}~\bibnamefont {Park}}, \bibinfo {author}
  {\bibfnamefont {N.~L.}\ \bibnamefont {Rodd}}, \bibinfo {author}
  {\bibfnamefont {K.}~\bibnamefont {Cranmer}},\ and\ \bibinfo {author}
  {\bibfnamefont {B.~R.}\ \bibnamefont {Safdi}},\ }\bibfield  {title} {\bibinfo
  {title} {{Deep Search for Decaying Dark Matter with XMM-Newton Blank-Sky
  Observations}},\ }\href {https://doi.org/10.1103/PhysRevLett.127.051101}
  {\bibfield  {journal} {\bibinfo  {journal} {Phys. Rev. Lett.}\ }\textbf
  {\bibinfo {volume} {127}},\ \bibinfo {pages} {051101} (\bibinfo {year}
  {2021})},\ \Eprint {https://arxiv.org/abs/2102.02207} {arXiv:2102.02207
  [astro-ph.CO]} \BibitemShut {NoStop}%
\bibitem [{\citenamefont {Roach}\ \emph {et~al.}(2020)\citenamefont {Roach},
  \citenamefont {Ng}, \citenamefont {Perez}, \citenamefont {Beacom},
  \citenamefont {Horiuchi}, \citenamefont {Krivonos},\ and\ \citenamefont
  {Wik}}]{Roach:2019ctw}%
  \BibitemOpen
  \bibfield  {author} {\bibinfo {author} {\bibfnamefont {B.~M.}\ \bibnamefont
  {Roach}}, \bibinfo {author} {\bibfnamefont {K.~C.~Y.}\ \bibnamefont {Ng}},
  \bibinfo {author} {\bibfnamefont {K.}~\bibnamefont {Perez}}, \bibinfo
  {author} {\bibfnamefont {J.~F.}\ \bibnamefont {Beacom}}, \bibinfo {author}
  {\bibfnamefont {S.}~\bibnamefont {Horiuchi}}, \bibinfo {author}
  {\bibfnamefont {R.}~\bibnamefont {Krivonos}},\ and\ \bibinfo {author}
  {\bibfnamefont {D.~R.}\ \bibnamefont {Wik}},\ }\bibfield  {title} {\bibinfo
  {title} {{NuSTAR Tests of Sterile-Neutrino Dark Matter: New Galactic Bulge
  Observations and Combined Impact}},\ }\href
  {https://doi.org/10.1103/PhysRevD.101.103011} {\bibfield  {journal} {\bibinfo
   {journal} {Phys. Rev. D}\ }\textbf {\bibinfo {volume} {101}},\ \bibinfo
  {pages} {103011} (\bibinfo {year} {2020})},\ \Eprint
  {https://arxiv.org/abs/1908.09037} {arXiv:1908.09037 [astro-ph.HE]}
  \BibitemShut {NoStop}%
\bibitem [{\citenamefont {Roach}\ \emph {et~al.}(2023)\citenamefont {Roach},
  \citenamefont {Rossland}, \citenamefont {Ng}, \citenamefont {Perez},
  \citenamefont {Beacom}, \citenamefont {Grefenstette}, \citenamefont
  {Horiuchi}, \citenamefont {Krivonos},\ and\ \citenamefont
  {Wik}}]{Roach:2022lgo}%
  \BibitemOpen
  \bibfield  {author} {\bibinfo {author} {\bibfnamefont {B.~M.}\ \bibnamefont
  {Roach}}, \bibinfo {author} {\bibfnamefont {S.}~\bibnamefont {Rossland}},
  \bibinfo {author} {\bibfnamefont {K.~C.~Y.}\ \bibnamefont {Ng}}, \bibinfo
  {author} {\bibfnamefont {K.}~\bibnamefont {Perez}}, \bibinfo {author}
  {\bibfnamefont {J.~F.}\ \bibnamefont {Beacom}}, \bibinfo {author}
  {\bibfnamefont {B.~W.}\ \bibnamefont {Grefenstette}}, \bibinfo {author}
  {\bibfnamefont {S.}~\bibnamefont {Horiuchi}}, \bibinfo {author}
  {\bibfnamefont {R.}~\bibnamefont {Krivonos}},\ and\ \bibinfo {author}
  {\bibfnamefont {D.~R.}\ \bibnamefont {Wik}},\ }\bibfield  {title} {\bibinfo
  {title} {{Long-exposure NuSTAR constraints on decaying dark matter in the
  Galactic halo}},\ }\href {https://doi.org/10.1103/PhysRevD.107.023009}
  {\bibfield  {journal} {\bibinfo  {journal} {Phys. Rev. D}\ }\textbf {\bibinfo
  {volume} {107}},\ \bibinfo {pages} {023009} (\bibinfo {year} {2023})},\
  \Eprint {https://arxiv.org/abs/2207.04572} {arXiv:2207.04572 [astro-ph.HE]}
  \BibitemShut {NoStop}%
\bibitem [{XRI(2020)}]{XRISMScienceTeam:2020rvx}%
  \BibitemOpen
  \bibfield  {title} {\bibinfo {title} {{Science with the X-ray Imaging and
  Spectroscopy Mission (XRISM)}},\ }\href@noop {} {\  (\bibinfo {year}
  {2020})},\ \Eprint {https://arxiv.org/abs/2003.04962} {arXiv:2003.04962
  [astro-ph.HE]} \BibitemShut {NoStop}%
\bibitem [{\citenamefont {Dessert}\ \emph
  {et~al.}(2020{\natexlab{b}})\citenamefont {Dessert}, \citenamefont {Rodd},\
  and\ \citenamefont {Safdi}}]{Dessert:2020hro}%
  \BibitemOpen
  \bibfield  {author} {\bibinfo {author} {\bibfnamefont {C.}~\bibnamefont
  {Dessert}}, \bibinfo {author} {\bibfnamefont {N.~L.}\ \bibnamefont {Rodd}},\
  and\ \bibinfo {author} {\bibfnamefont {B.~R.}\ \bibnamefont {Safdi}},\
  }\bibfield  {title} {\bibinfo {title} {{Response to a comment on Dessert et
  al. \textquotedblleft{}The dark matter interpretation of the 3.5 keV line is
  inconsistent with blank-sky observations\textquotedblright{}}},\ }\href
  {https://doi.org/10.1016/j.dark.2020.100656} {\bibfield  {journal} {\bibinfo
  {journal} {Phys. Dark Univ.}\ }\textbf {\bibinfo {volume} {30}},\ \bibinfo
  {pages} {100656} (\bibinfo {year} {2020}{\natexlab{b}})},\ \Eprint
  {https://arxiv.org/abs/2006.03974} {arXiv:2006.03974 [astro-ph.CO]}
  \BibitemShut {NoStop}%
\bibitem [{Note1()}]{Note1}%
  \BibitemOpen
  \bibinfo {note} {The 3.5 keV line is an observed unassociated X-ray line
  found using a variety of instruments, including XMM-Newton and Chandra, from
  observations of a number of targets, including the Perseus galaxy cluster,
  nearby galaxies such as M31, and blank regions of the Milky Way~\cite
  {Bulbul:2014sua,Boyarsky:2014jta,Urban:2014yda,Jeltema:2014qfa,Cappelluti:2017ywp}.
  Nonetheless, Refs.~\cite {Dessert:2018qih,Dessert:2020hro,Foster:2021ngm}
  found no evidence for the line and were able to exclude decaying DM as an
  explanation.}\BibitemShut {Stop}%
\bibitem [{\citenamefont {Abazajian}(2017)}]{Abazajian:2017tcc}%
  \BibitemOpen
  \bibfield  {author} {\bibinfo {author} {\bibfnamefont {K.~N.}\ \bibnamefont
  {Abazajian}},\ }\bibfield  {title} {\bibinfo {title} {{Sterile neutrinos in
  cosmology}},\ }\href {https://doi.org/10.1016/j.physrep.2017.10.003}
  {\bibfield  {journal} {\bibinfo  {journal} {Phys. Rept.}\ }\textbf {\bibinfo
  {volume} {711-712}},\ \bibinfo {pages} {1} (\bibinfo {year} {2017})},\
  \Eprint {https://arxiv.org/abs/1705.01837} {arXiv:1705.01837 [hep-ph]}
  \BibitemShut {NoStop}%
\bibitem [{\citenamefont {Boyarsky}\ \emph {et~al.}(2019)\citenamefont
  {Boyarsky}, \citenamefont {Drewes}, \citenamefont {Lasserre}, \citenamefont
  {Mertens},\ and\ \citenamefont {Ruchayskiy}}]{Boyarsky:2018tvu}%
  \BibitemOpen
  \bibfield  {author} {\bibinfo {author} {\bibfnamefont {A.}~\bibnamefont
  {Boyarsky}}, \bibinfo {author} {\bibfnamefont {M.}~\bibnamefont {Drewes}},
  \bibinfo {author} {\bibfnamefont {T.}~\bibnamefont {Lasserre}}, \bibinfo
  {author} {\bibfnamefont {S.}~\bibnamefont {Mertens}},\ and\ \bibinfo {author}
  {\bibfnamefont {O.}~\bibnamefont {Ruchayskiy}},\ }\bibfield  {title}
  {\bibinfo {title} {{Sterile neutrino Dark Matter}},\ }\href
  {https://doi.org/10.1016/j.ppnp.2018.07.004} {\bibfield  {journal} {\bibinfo
  {journal} {Prog. Part. Nucl. Phys.}\ }\textbf {\bibinfo {volume} {104}},\
  \bibinfo {pages} {1} (\bibinfo {year} {2019})},\ \Eprint
  {https://arxiv.org/abs/1807.07938} {arXiv:1807.07938 [hep-ph]} \BibitemShut
  {NoStop}%
\bibitem [{\citenamefont {Dasgupta}\ and\ \citenamefont
  {Kopp}(2021)}]{Dasgupta:2021ies}%
  \BibitemOpen
  \bibfield  {author} {\bibinfo {author} {\bibfnamefont {B.}~\bibnamefont
  {Dasgupta}}\ and\ \bibinfo {author} {\bibfnamefont {J.}~\bibnamefont
  {Kopp}},\ }\bibfield  {title} {\bibinfo {title} {{Sterile Neutrinos}},\
  }\href {https://doi.org/10.1016/j.physrep.2021.06.002} {\bibfield  {journal}
  {\bibinfo  {journal} {Phys. Rept.}\ }\textbf {\bibinfo {volume} {928}},\
  \bibinfo {pages} {1} (\bibinfo {year} {2021})},\ \Eprint
  {https://arxiv.org/abs/2106.05913} {arXiv:2106.05913 [hep-ph]} \BibitemShut
  {NoStop}%
\bibitem [{\citenamefont {Pal}\ and\ \citenamefont
  {Wolfenstein}(1982)}]{Pal:1981rm}%
  \BibitemOpen
  \bibfield  {author} {\bibinfo {author} {\bibfnamefont {P.~B.}\ \bibnamefont
  {Pal}}\ and\ \bibinfo {author} {\bibfnamefont {L.}~\bibnamefont
  {Wolfenstein}},\ }\bibfield  {title} {\bibinfo {title} {{Radiative Decays of
  Massive Neutrinos}},\ }\href {https://doi.org/10.1103/PhysRevD.25.766}
  {\bibfield  {journal} {\bibinfo  {journal} {Phys. Rev. D}\ }\textbf {\bibinfo
  {volume} {25}},\ \bibinfo {pages} {766} (\bibinfo {year} {1982})}\BibitemShut
  {NoStop}%
\bibitem [{\citenamefont {Cherry}\ and\ \citenamefont
  {Horiuchi}(2017)}]{Cherry:2017dwu}%
  \BibitemOpen
  \bibfield  {author} {\bibinfo {author} {\bibfnamefont {J.~F.}\ \bibnamefont
  {Cherry}}\ and\ \bibinfo {author} {\bibfnamefont {S.}~\bibnamefont
  {Horiuchi}},\ }\bibfield  {title} {\bibinfo {title} {{Closing in on
  Resonantly Produced Sterile Neutrino Dark Matter}},\ }\href
  {https://doi.org/10.1103/PhysRevD.95.083015} {\bibfield  {journal} {\bibinfo
  {journal} {Phys. Rev. D}\ }\textbf {\bibinfo {volume} {95}},\ \bibinfo
  {pages} {083015} (\bibinfo {year} {2017})},\ \Eprint
  {https://arxiv.org/abs/1701.07874} {arXiv:1701.07874 [hep-ph]} \BibitemShut
  {NoStop}%
\bibitem [{\citenamefont {Nadler}\ \emph {et~al.}(2021)\citenamefont {Nadler}
  \emph {et~al.}}]{DES:2020fxi}%
  \BibitemOpen
  \bibfield  {author} {\bibinfo {author} {\bibfnamefont {E.~O.}\ \bibnamefont
  {Nadler}} \emph {et~al.} (\bibinfo {collaboration} {DES}),\ }\bibfield
  {title} {\bibinfo {title} {{Milky Way Satellite Census. III. Constraints on
  Dark Matter Properties from Observations of Milky Way Satellite Galaxies}},\
  }\href {https://doi.org/10.1103/PhysRevLett.126.091101} {\bibfield  {journal}
  {\bibinfo  {journal} {Phys. Rev. Lett.}\ }\textbf {\bibinfo {volume} {126}},\
  \bibinfo {pages} {091101} (\bibinfo {year} {2021})},\ \Eprint
  {https://arxiv.org/abs/2008.00022} {arXiv:2008.00022 [astro-ph.CO]}
  \BibitemShut {NoStop}%
\bibitem [{\citenamefont {Dodelson}\ and\ \citenamefont
  {Widrow}(1994)}]{Dodelson:1993je}%
  \BibitemOpen
  \bibfield  {author} {\bibinfo {author} {\bibfnamefont {S.}~\bibnamefont
  {Dodelson}}\ and\ \bibinfo {author} {\bibfnamefont {L.~M.}\ \bibnamefont
  {Widrow}},\ }\bibfield  {title} {\bibinfo {title} {{Sterile-neutrinos as dark
  matter}},\ }\href {https://doi.org/10.1103/PhysRevLett.72.17} {\bibfield
  {journal} {\bibinfo  {journal} {Phys. Rev. Lett.}\ }\textbf {\bibinfo
  {volume} {72}},\ \bibinfo {pages} {17} (\bibinfo {year} {1994})},\ \Eprint
  {https://arxiv.org/abs/hep-ph/9303287} {arXiv:hep-ph/9303287 [hep-ph]}
  \BibitemShut {NoStop}%
\bibitem [{\citenamefont {Shi}\ and\ \citenamefont
  {Fuller}(1999)}]{Shi:1998km}%
  \BibitemOpen
  \bibfield  {author} {\bibinfo {author} {\bibfnamefont {X.-D.}\ \bibnamefont
  {Shi}}\ and\ \bibinfo {author} {\bibfnamefont {G.~M.}\ \bibnamefont
  {Fuller}},\ }\bibfield  {title} {\bibinfo {title} {{A New dark matter
  candidate: Nonthermal sterile neutrinos}},\ }\href
  {https://doi.org/10.1103/PhysRevLett.82.2832} {\bibfield  {journal} {\bibinfo
   {journal} {Phys. Rev. Lett.}\ }\textbf {\bibinfo {volume} {82}},\ \bibinfo
  {pages} {2832} (\bibinfo {year} {1999})},\ \Eprint
  {https://arxiv.org/abs/astro-ph/9810076} {arXiv:astro-ph/9810076 [astro-ph]}
  \BibitemShut {NoStop}%
\bibitem [{\citenamefont {An}\ \emph {et~al.}(2023)\citenamefont {An},
  \citenamefont {Gluscevic}, \citenamefont {Nadler},\ and\ \citenamefont
  {Zhang}}]{An:2023mkf}%
  \BibitemOpen
  \bibfield  {author} {\bibinfo {author} {\bibfnamefont {R.}~\bibnamefont
  {An}}, \bibinfo {author} {\bibfnamefont {V.}~\bibnamefont {Gluscevic}},
  \bibinfo {author} {\bibfnamefont {E.~O.}\ \bibnamefont {Nadler}},\ and\
  \bibinfo {author} {\bibfnamefont {Y.}~\bibnamefont {Zhang}},\ }\bibfield
  {title} {\bibinfo {title} {{Can Neutrino Self-interactions Save Sterile
  Neutrino Dark Matter?}},\ }\href@noop {} {\  (\bibinfo {year} {2023})},\
  \Eprint {https://arxiv.org/abs/2301.08299} {arXiv:2301.08299 [astro-ph.CO]}
  \BibitemShut {NoStop}%
\bibitem [{\citenamefont {De~Gouv\^ea}\ \emph {et~al.}(2020)\citenamefont
  {De~Gouv\^ea}, \citenamefont {Sen}, \citenamefont {Tangarife},\ and\
  \citenamefont {Zhang}}]{DeGouvea:2019wpf}%
  \BibitemOpen
  \bibfield  {author} {\bibinfo {author} {\bibfnamefont {A.}~\bibnamefont
  {De~Gouv\^ea}}, \bibinfo {author} {\bibfnamefont {M.}~\bibnamefont {Sen}},
  \bibinfo {author} {\bibfnamefont {W.}~\bibnamefont {Tangarife}},\ and\
  \bibinfo {author} {\bibfnamefont {Y.}~\bibnamefont {Zhang}},\ }\bibfield
  {title} {\bibinfo {title} {{Dodelson-Widrow Mechanism in the Presence of
  Self-Interacting Neutrinos}},\ }\href
  {https://doi.org/10.1103/PhysRevLett.124.081802} {\bibfield  {journal}
  {\bibinfo  {journal} {Phys. Rev. Lett.}\ }\textbf {\bibinfo {volume} {124}},\
  \bibinfo {pages} {081802} (\bibinfo {year} {2020})},\ \Eprint
  {https://arxiv.org/abs/1910.04901} {arXiv:1910.04901 [hep-ph]} \BibitemShut
  {NoStop}%
\bibitem [{\citenamefont {Bringmann}\ \emph {et~al.}(2023)\citenamefont
  {Bringmann}, \citenamefont {Depta}, \citenamefont {Hufnagel}, \citenamefont
  {Kersten}, \citenamefont {Ruderman},\ and\ \citenamefont
  {Schmidt-Hoberg}}]{Bringmann:2022aim}%
  \BibitemOpen
  \bibfield  {author} {\bibinfo {author} {\bibfnamefont {T.}~\bibnamefont
  {Bringmann}}, \bibinfo {author} {\bibfnamefont {P.~F.}\ \bibnamefont
  {Depta}}, \bibinfo {author} {\bibfnamefont {M.}~\bibnamefont {Hufnagel}},
  \bibinfo {author} {\bibfnamefont {J.}~\bibnamefont {Kersten}}, \bibinfo
  {author} {\bibfnamefont {J.~T.}\ \bibnamefont {Ruderman}},\ and\ \bibinfo
  {author} {\bibfnamefont {K.}~\bibnamefont {Schmidt-Hoberg}},\ }\bibfield
  {title} {\bibinfo {title} {{Minimal sterile neutrino dark matter}},\ }\href
  {https://doi.org/10.1103/PhysRevD.107.L071702} {\bibfield  {journal}
  {\bibinfo  {journal} {Phys. Rev. D}\ }\textbf {\bibinfo {volume} {107}},\
  \bibinfo {pages} {L071702} (\bibinfo {year} {2023})},\ \Eprint
  {https://arxiv.org/abs/2206.10630} {arXiv:2206.10630 [hep-ph]} \BibitemShut
  {NoStop}%
\bibitem [{\citenamefont {Higaki}\ \emph {et~al.}(2014)\citenamefont {Higaki},
  \citenamefont {Jeong},\ and\ \citenamefont {Takahashi}}]{Higaki:2014zua}%
  \BibitemOpen
  \bibfield  {author} {\bibinfo {author} {\bibfnamefont {T.}~\bibnamefont
  {Higaki}}, \bibinfo {author} {\bibfnamefont {K.~S.}\ \bibnamefont {Jeong}},\
  and\ \bibinfo {author} {\bibfnamefont {F.}~\bibnamefont {Takahashi}},\
  }\bibfield  {title} {\bibinfo {title} {{The 7 keV axion dark matter and the
  X-ray line signal}},\ }\href {https://doi.org/10.1016/j.physletb.2014.04.007}
  {\bibfield  {journal} {\bibinfo  {journal} {Phys. Lett. B}\ }\textbf
  {\bibinfo {volume} {733}},\ \bibinfo {pages} {25} (\bibinfo {year} {2014})},\
  \Eprint {https://arxiv.org/abs/1402.6965} {arXiv:1402.6965 [hep-ph]}
  \BibitemShut {NoStop}%
\bibitem [{\citenamefont {Jaeckel}\ \emph {et~al.}(2014)\citenamefont
  {Jaeckel}, \citenamefont {Redondo},\ and\ \citenamefont
  {Ringwald}}]{Jaeckel:2014qea}%
  \BibitemOpen
  \bibfield  {author} {\bibinfo {author} {\bibfnamefont {J.}~\bibnamefont
  {Jaeckel}}, \bibinfo {author} {\bibfnamefont {J.}~\bibnamefont {Redondo}},\
  and\ \bibinfo {author} {\bibfnamefont {A.}~\bibnamefont {Ringwald}},\
  }\bibfield  {title} {\bibinfo {title} {{3.55 keV hint for decaying axionlike
  particle dark matter}},\ }\href {https://doi.org/10.1103/PhysRevD.89.103511}
  {\bibfield  {journal} {\bibinfo  {journal} {Phys. Rev. D}\ }\textbf {\bibinfo
  {volume} {89}},\ \bibinfo {pages} {103511} (\bibinfo {year} {2014})},\
  \Eprint {https://arxiv.org/abs/1402.7335} {arXiv:1402.7335 [hep-ph]}
  \BibitemShut {NoStop}%
\bibitem [{\citenamefont {Foster}\ \emph {et~al.}(2022)\citenamefont {Foster},
  \citenamefont {Kumar}, \citenamefont {Safdi},\ and\ \citenamefont
  {Soreq}}]{Foster:2022ajl}%
  \BibitemOpen
  \bibfield  {author} {\bibinfo {author} {\bibfnamefont {J.~W.}\ \bibnamefont
  {Foster}}, \bibinfo {author} {\bibfnamefont {S.}~\bibnamefont {Kumar}},
  \bibinfo {author} {\bibfnamefont {B.~R.}\ \bibnamefont {Safdi}},\ and\
  \bibinfo {author} {\bibfnamefont {Y.}~\bibnamefont {Soreq}},\ }\bibfield
  {title} {\bibinfo {title} {{Dark Grand Unification in the axiverse: decaying
  axion dark matter and spontaneous baryogenesis}},\ }\href
  {https://doi.org/10.1007/JHEP12(2022)119} {\bibfield  {journal} {\bibinfo
  {journal} {JHEP}\ }\textbf {\bibinfo {volume} {12}},\ \bibinfo {pages}
  {119}},\ \Eprint {https://arxiv.org/abs/2208.10504} {arXiv:2208.10504
  [hep-ph]} \BibitemShut {NoStop}%
\bibitem [{\citenamefont {Panci}\ \emph {et~al.}(2023)\citenamefont {Panci},
  \citenamefont {Redigolo}, \citenamefont {Schwetz},\ and\ \citenamefont
  {Ziegler}}]{Panci:2022wlc}%
  \BibitemOpen
  \bibfield  {author} {\bibinfo {author} {\bibfnamefont {P.}~\bibnamefont
  {Panci}}, \bibinfo {author} {\bibfnamefont {D.}~\bibnamefont {Redigolo}},
  \bibinfo {author} {\bibfnamefont {T.}~\bibnamefont {Schwetz}},\ and\ \bibinfo
  {author} {\bibfnamefont {R.}~\bibnamefont {Ziegler}},\ }\bibfield  {title}
  {\bibinfo {title} {{Axion dark matter from lepton flavor-violating decays}},\
  }\href {https://doi.org/10.1016/j.physletb.2023.137919} {\bibfield  {journal}
  {\bibinfo  {journal} {Phys. Lett. B}\ }\textbf {\bibinfo {volume} {841}},\
  \bibinfo {pages} {137919} (\bibinfo {year} {2023})},\ \Eprint
  {https://arxiv.org/abs/2209.03371} {arXiv:2209.03371 [hep-ph]} \BibitemShut
  {NoStop}%
\bibitem [{\citenamefont {Langhoff}\ \emph {et~al.}(2022)\citenamefont
  {Langhoff}, \citenamefont {Outmezguine},\ and\ \citenamefont
  {Rodd}}]{Langhoff:2022bij}%
  \BibitemOpen
  \bibfield  {author} {\bibinfo {author} {\bibfnamefont {K.}~\bibnamefont
  {Langhoff}}, \bibinfo {author} {\bibfnamefont {N.~J.}\ \bibnamefont
  {Outmezguine}},\ and\ \bibinfo {author} {\bibfnamefont {N.~L.}\ \bibnamefont
  {Rodd}},\ }\bibfield  {title} {\bibinfo {title} {{Irreducible Axion
  Background}},\ }\href {https://doi.org/10.1103/PhysRevLett.129.241101}
  {\bibfield  {journal} {\bibinfo  {journal} {Phys. Rev. Lett.}\ }\textbf
  {\bibinfo {volume} {129}},\ \bibinfo {pages} {241101} (\bibinfo {year}
  {2022})},\ \Eprint {https://arxiv.org/abs/2209.06216} {arXiv:2209.06216
  [hep-ph]} \BibitemShut {NoStop}%
\bibitem [{\citenamefont {Speckhard}\ \emph {et~al.}(2016)\citenamefont
  {Speckhard}, \citenamefont {Ng}, \citenamefont {Beacom},\ and\ \citenamefont
  {Laha}}]{Speckhard:2015eva}%
  \BibitemOpen
  \bibfield  {author} {\bibinfo {author} {\bibfnamefont {E.~G.}\ \bibnamefont
  {Speckhard}}, \bibinfo {author} {\bibfnamefont {K.~C.~Y.}\ \bibnamefont
  {Ng}}, \bibinfo {author} {\bibfnamefont {J.~F.}\ \bibnamefont {Beacom}},\
  and\ \bibinfo {author} {\bibfnamefont {R.}~\bibnamefont {Laha}},\ }\bibfield
  {title} {\bibinfo {title} {{Dark Matter Velocity Spectroscopy}},\ }\href
  {https://doi.org/10.1103/PhysRevLett.116.031301} {\bibfield  {journal}
  {\bibinfo  {journal} {Phys. Rev. Lett.}\ }\textbf {\bibinfo {volume} {116}},\
  \bibinfo {pages} {031301} (\bibinfo {year} {2016})},\ \Eprint
  {https://arxiv.org/abs/1507.04744} {arXiv:1507.04744 [astro-ph.CO]}
  \BibitemShut {NoStop}%
\bibitem [{\citenamefont {Schoenrich}\ \emph {et~al.}(2010)\citenamefont
  {Schoenrich}, \citenamefont {Binney},\ and\ \citenamefont
  {Dehnen}}]{Schoenrich:2009bx}%
  \BibitemOpen
  \bibfield  {author} {\bibinfo {author} {\bibfnamefont {R.}~\bibnamefont
  {Schoenrich}}, \bibinfo {author} {\bibfnamefont {J.}~\bibnamefont {Binney}},\
  and\ \bibinfo {author} {\bibfnamefont {W.}~\bibnamefont {Dehnen}},\
  }\bibfield  {title} {\bibinfo {title} {{Local Kinematics and the Local
  Standard of Rest}},\ }\href
  {https://doi.org/10.1111/j.1365-2966.2010.16253.x} {\bibfield  {journal}
  {\bibinfo  {journal} {Mon. Not. Roy. Astron. Soc.}\ }\textbf {\bibinfo
  {volume} {403}},\ \bibinfo {pages} {1829} (\bibinfo {year} {2010})},\ \Eprint
  {https://arxiv.org/abs/0912.3693} {arXiv:0912.3693 [astro-ph.GA]}
  \BibitemShut {NoStop}%
\bibitem [{\citenamefont {{Mignard}}(2000)}]{2000A&A...354..522M}%
  \BibitemOpen
  \bibfield  {author} {\bibinfo {author} {\bibfnamefont {F.}~\bibnamefont
  {{Mignard}}},\ }\bibfield  {title} {\bibinfo {title} {{Local galactic
  kinematics from Hipparcos proper motions}},\ }\href@noop {} {\bibfield
  {journal} {\bibinfo  {journal} {Astron. Astrophys.}\ }\textbf {\bibinfo
  {volume} {354}},\ \bibinfo {pages} {522} (\bibinfo {year}
  {2000})}\BibitemShut {NoStop}%
\bibitem [{\citenamefont {Navarro}\ \emph {et~al.}(1996)\citenamefont
  {Navarro}, \citenamefont {Frenk},\ and\ \citenamefont
  {White}}]{Navarro:1995iw}%
  \BibitemOpen
  \bibfield  {author} {\bibinfo {author} {\bibfnamefont {J.~F.}\ \bibnamefont
  {Navarro}}, \bibinfo {author} {\bibfnamefont {C.~S.}\ \bibnamefont {Frenk}},\
  and\ \bibinfo {author} {\bibfnamefont {S.~D.~M.}\ \bibnamefont {White}},\
  }\bibfield  {title} {\bibinfo {title} {{The Structure of cold dark matter
  halos}},\ }\href {https://doi.org/10.1086/177173} {\bibfield  {journal}
  {\bibinfo  {journal} {Astrophys. J.}\ }\textbf {\bibinfo {volume} {462}},\
  \bibinfo {pages} {563} (\bibinfo {year} {1996})},\ \Eprint
  {https://arxiv.org/abs/astro-ph/9508025} {astro-ph/9508025} \BibitemShut
  {NoStop}%
\bibitem [{\citenamefont {Navarro}\ \emph {et~al.}(1997)\citenamefont
  {Navarro}, \citenamefont {Frenk},\ and\ \citenamefont
  {White}}]{Navarro:1996gj}%
  \BibitemOpen
  \bibfield  {author} {\bibinfo {author} {\bibfnamefont {J.~F.}\ \bibnamefont
  {Navarro}}, \bibinfo {author} {\bibfnamefont {C.~S.}\ \bibnamefont {Frenk}},\
  and\ \bibinfo {author} {\bibfnamefont {S.~D.~M.}\ \bibnamefont {White}},\
  }\bibfield  {title} {\bibinfo {title} {{A Universal density profile from
  hierarchical clustering}},\ }\href {https://doi.org/10.1086/304888}
  {\bibfield  {journal} {\bibinfo  {journal} {Astrophys. J.}\ }\textbf
  {\bibinfo {volume} {490}},\ \bibinfo {pages} {493} (\bibinfo {year}
  {1997})},\ \Eprint {https://arxiv.org/abs/astro-ph/9611107}
  {arXiv:astro-ph/9611107 [astro-ph]} \BibitemShut {NoStop}%
\bibitem [{\citenamefont {Cautun}\ \emph {et~al.}(2020)\citenamefont {Cautun},
  \citenamefont {Benitez-Llambay}, \citenamefont {Deason}, \citenamefont
  {Frenk}, \citenamefont {Fattahi}, \citenamefont {G\'omez}, \citenamefont
  {Grand}, \citenamefont {Oman}, \citenamefont {Navarro},\ and\ \citenamefont
  {Simpson}}]{Cautun:2019eaf}%
  \BibitemOpen
  \bibfield  {author} {\bibinfo {author} {\bibfnamefont {M.}~\bibnamefont
  {Cautun}}, \bibinfo {author} {\bibfnamefont {A.}~\bibnamefont
  {Benitez-Llambay}}, \bibinfo {author} {\bibfnamefont {A.~J.}\ \bibnamefont
  {Deason}}, \bibinfo {author} {\bibfnamefont {C.~S.}\ \bibnamefont {Frenk}},
  \bibinfo {author} {\bibfnamefont {A.}~\bibnamefont {Fattahi}}, \bibinfo
  {author} {\bibfnamefont {F.~A.}\ \bibnamefont {G\'omez}}, \bibinfo {author}
  {\bibfnamefont {R.~J.~J.}\ \bibnamefont {Grand}}, \bibinfo {author}
  {\bibfnamefont {K.~A.}\ \bibnamefont {Oman}}, \bibinfo {author}
  {\bibfnamefont {J.~F.}\ \bibnamefont {Navarro}},\ and\ \bibinfo {author}
  {\bibfnamefont {C.~M.}\ \bibnamefont {Simpson}},\ }\bibfield  {title}
  {\bibinfo {title} {{The Milky Way total mass profile as inferred from Gaia
  DR2}},\ }\href {https://doi.org/10.1093/mnras/staa1017} {\bibfield  {journal}
  {\bibinfo  {journal} {Mon. Not. Roy. Astron. Soc.}\ }\textbf {\bibinfo
  {volume} {494}},\ \bibinfo {pages} {4291} (\bibinfo {year} {2020})},\ \Eprint
  {https://arxiv.org/abs/1911.04557} {arXiv:1911.04557 [astro-ph.GA]}
  \BibitemShut {NoStop}%
\bibitem [{\citenamefont {Cowan}\ \emph
  {et~al.}(2011{\natexlab{a}})\citenamefont {Cowan}, \citenamefont {Cranmer},
  \citenamefont {Gross},\ and\ \citenamefont {Vitells}}]{Cowan:2010js}%
  \BibitemOpen
  \bibfield  {author} {\bibinfo {author} {\bibfnamefont {G.}~\bibnamefont
  {Cowan}}, \bibinfo {author} {\bibfnamefont {K.}~\bibnamefont {Cranmer}},
  \bibinfo {author} {\bibfnamefont {E.}~\bibnamefont {Gross}},\ and\ \bibinfo
  {author} {\bibfnamefont {O.}~\bibnamefont {Vitells}},\ }\bibfield  {title}
  {\bibinfo {title} {{Asymptotic formulae for likelihood-based tests of new
  physics}},\ }\href {https://doi.org/10.1140/epjc/s10052-011-1554-0,
  10.1140/epjc/s10052-013-2501-z} {\bibfield  {journal} {\bibinfo  {journal}
  {Eur. Phys. J.}\ }\textbf {\bibinfo {volume} {C71}},\ \bibinfo {pages} {1554}
  (\bibinfo {year} {2011}{\natexlab{a}})},\ \bibinfo {note} {[Erratum: Eur.
  Phys. J.C73,2501(2013)]},\ \Eprint {https://arxiv.org/abs/1007.1727}
  {arXiv:1007.1727 [physics.data-an]} \BibitemShut {NoStop}%
\bibitem [{\citenamefont {Cowan}\ \emph
  {et~al.}(2011{\natexlab{b}})\citenamefont {Cowan}, \citenamefont {Cranmer},
  \citenamefont {Gross},\ and\ \citenamefont {Vitells}}]{Cowan:2011an}%
  \BibitemOpen
  \bibfield  {author} {\bibinfo {author} {\bibfnamefont {G.}~\bibnamefont
  {Cowan}}, \bibinfo {author} {\bibfnamefont {K.}~\bibnamefont {Cranmer}},
  \bibinfo {author} {\bibfnamefont {E.}~\bibnamefont {Gross}},\ and\ \bibinfo
  {author} {\bibfnamefont {O.}~\bibnamefont {Vitells}},\ }\bibfield  {title}
  {\bibinfo {title} {{Power-Constrained Limits}},\ }\href@noop {} {\  (\bibinfo
  {year} {2011}{\natexlab{b}})},\ \Eprint {https://arxiv.org/abs/1105.3166}
  {arXiv:1105.3166 [physics.data-an]} \BibitemShut {NoStop}%
\bibitem [{sup()}]{supp_data}%
  \BibitemOpen
  \bibinfo {note} {See
  \href{https://github.com/bsafdi/Hitomi_BSO_for_DM}{github.com/bsafdi/Hitomi\_BSO\_for\_DM}}\BibitemShut
  {NoStop}%
\bibitem [{\citenamefont {Deslattes}\ \emph {et~al.}(2003)\citenamefont
  {Deslattes}, \citenamefont {Kessler}, \citenamefont {Indelicato},
  \citenamefont {de~Billy}, \citenamefont {Lindroth},\ and\ \citenamefont
  {Anton}}]{RevModPhys.75.35}%
  \BibitemOpen
  \bibfield  {author} {\bibinfo {author} {\bibfnamefont {R.~D.}\ \bibnamefont
  {Deslattes}}, \bibinfo {author} {\bibfnamefont {E.~G.}\ \bibnamefont
  {Kessler}}, \bibinfo {author} {\bibfnamefont {P.}~\bibnamefont {Indelicato}},
  \bibinfo {author} {\bibfnamefont {L.}~\bibnamefont {de~Billy}}, \bibinfo
  {author} {\bibfnamefont {E.}~\bibnamefont {Lindroth}},\ and\ \bibinfo
  {author} {\bibfnamefont {J.}~\bibnamefont {Anton}},\ }\bibfield  {title}
  {\bibinfo {title} {X-ray transition energies: new approach to a comprehensive
  evaluation},\ }\href {https://doi.org/10.1103/RevModPhys.75.35} {\bibfield
  {journal} {\bibinfo  {journal} {Rev. Mod. Phys.}\ }\textbf {\bibinfo {volume}
  {75}},\ \bibinfo {pages} {35} (\bibinfo {year} {2003})}\BibitemShut {NoStop}%
\bibitem [{\citenamefont {Aharonian}\ \emph {et~al.}(2018)\citenamefont
  {Aharonian} \emph {et~al.}}]{Hitomi:2017qip}%
  \BibitemOpen
  \bibfield  {author} {\bibinfo {author} {\bibfnamefont {F.}~\bibnamefont
  {Aharonian}} \emph {et~al.} (\bibinfo {collaboration} {Hitomi}),\ }\bibfield
  {title} {\bibinfo {title} {{Glimpse of the highly obscured HMXB IGR
  J16318\ensuremath{-}4848 with Hitomi}},\ }\href
  {https://doi.org/10.1093/pasj/psx154} {\bibfield  {journal} {\bibinfo
  {journal} {Publ. Astron. Soc. Jap.}\ }\textbf {\bibinfo {volume} {70}},\
  \bibinfo {pages} {17} (\bibinfo {year} {2018})},\ \Eprint
  {https://arxiv.org/abs/1711.07727} {arXiv:1711.07727 [astro-ph.HE]}
  \BibitemShut {NoStop}%
\bibitem [{\citenamefont {Webb}\ \emph {et~al.}(2020)\citenamefont {Webb} \emph
  {et~al.}}]{Webb:2020rgy}%
  \BibitemOpen
  \bibfield  {author} {\bibinfo {author} {\bibfnamefont {N.~A.}\ \bibnamefont
  {Webb}} \emph {et~al.},\ }\bibfield  {title} {\bibinfo {title} {{The
  XMM-Newton serendipitous survey IX. The fourth XMM-Newton serendipitous
  source catalogue}},\ }\href {https://doi.org/10.1051/0004-6361/201937353}
  {\bibfield  {journal} {\bibinfo  {journal} {Astron. Astrophys.}\ }\textbf
  {\bibinfo {volume} {641}},\ \bibinfo {pages} {A136} (\bibinfo {year}
  {2020})},\ \Eprint {https://arxiv.org/abs/2007.02899} {arXiv:2007.02899
  [astro-ph.HE]} \BibitemShut {NoStop}%
\bibitem [{\citenamefont {Ando}\ \emph {et~al.}(2021)\citenamefont {Ando} \emph
  {et~al.}}]{Ando:2021fhj}%
  \BibitemOpen
  \bibfield  {author} {\bibinfo {author} {\bibfnamefont {S.}~\bibnamefont
  {Ando}} \emph {et~al.},\ }\bibfield  {title} {\bibinfo {title} {{Decaying
  dark matter in dwarf spheroidal galaxies: Prospects for x-ray and gamma-ray
  telescopes}},\ }\href {https://doi.org/10.1103/PhysRevD.104.023022}
  {\bibfield  {journal} {\bibinfo  {journal} {Phys. Rev. D}\ }\textbf {\bibinfo
  {volume} {104}},\ \bibinfo {pages} {023022} (\bibinfo {year} {2021})},\
  \Eprint {https://arxiv.org/abs/2103.13242} {arXiv:2103.13242 [astro-ph.HE]}
  \BibitemShut {NoStop}%
\bibitem [{\citenamefont {Lovell}\ \emph {et~al.}(2018)\citenamefont {Lovell}
  \emph {et~al.}}]{Lovell:2018xng}%
  \BibitemOpen
  \bibfield  {author} {\bibinfo {author} {\bibfnamefont {M.~R.}\ \bibnamefont
  {Lovell}} \emph {et~al.},\ }\bibfield  {title} {\bibinfo {title} {{The signal
  of decaying dark matter with hydrodynamical simulations}},\ }\href@noop {} {\
   (\bibinfo {year} {2018})},\ \Eprint {https://arxiv.org/abs/1810.05168}
  {arXiv:1810.05168 [astro-ph.CO]} \BibitemShut {NoStop}%
\bibitem [{\citenamefont {Lovell}\ \emph {et~al.}(2019)\citenamefont {Lovell},
  \citenamefont {Iakubovskyi}, \citenamefont {Barnes}, \citenamefont {Bose},
  \citenamefont {Frenk}, \citenamefont {Theuns},\ and\ \citenamefont
  {Hellwing}}]{Lovell:2019aol}%
  \BibitemOpen
  \bibfield  {author} {\bibinfo {author} {\bibfnamefont {M.~R.}\ \bibnamefont
  {Lovell}}, \bibinfo {author} {\bibfnamefont {D.}~\bibnamefont {Iakubovskyi}},
  \bibinfo {author} {\bibfnamefont {D.}~\bibnamefont {Barnes}}, \bibinfo
  {author} {\bibfnamefont {S.}~\bibnamefont {Bose}}, \bibinfo {author}
  {\bibfnamefont {C.~S.}\ \bibnamefont {Frenk}}, \bibinfo {author}
  {\bibfnamefont {T.}~\bibnamefont {Theuns}},\ and\ \bibinfo {author}
  {\bibfnamefont {W.~A.}\ \bibnamefont {Hellwing}},\ }\bibfield  {title}
  {\bibinfo {title} {{Simulating the dark matter decay signal from the Perseus
  galaxy cluster}},\ }\href {https://doi.org/10.3847/2041-8213/ab13ac}
  {\bibfield  {journal} {\bibinfo  {journal} {Astrophys. J. Lett.}\ }\textbf
  {\bibinfo {volume} {875}},\ \bibinfo {pages} {L24} (\bibinfo {year}
  {2019})},\ \Eprint {https://arxiv.org/abs/1903.11608} {arXiv:1903.11608
  [astro-ph.CO]} \BibitemShut {NoStop}%
\bibitem [{\citenamefont {Lovell}(2023)}]{Lovell:2023olv}%
  \BibitemOpen
  \bibfield  {author} {\bibinfo {author} {\bibfnamefont {M.~R.}\ \bibnamefont
  {Lovell}},\ }\bibfield  {title} {\bibinfo {title} {{Anticipating the XRISM
  search for the decay of resonantly produced sterile neutrino dark matter}},\
  }\href@noop {} {\  (\bibinfo {year} {2023})},\ \Eprint
  {https://arxiv.org/abs/2303.15513} {arXiv:2303.15513 [astro-ph.CO]}
  \BibitemShut {NoStop}%
\bibitem [{\citenamefont {Barret}\ \emph {et~al.}(2013)\citenamefont {Barret}
  \emph {et~al.}}]{Barret:2013mxa}%
  \BibitemOpen
  \bibfield  {author} {\bibinfo {author} {\bibfnamefont {D.}~\bibnamefont
  {Barret}} \emph {et~al.},\ }\bibfield  {title} {\bibinfo {title} {{The Hot
  and Energetic Universe: The X-ray Integral Field Unit (X-IFU) for Athena+}},\
  }\href@noop {} {\  (\bibinfo {year} {2013})},\ \Eprint
  {https://arxiv.org/abs/1308.6784} {arXiv:1308.6784 [astro-ph.IM]}
  \BibitemShut {NoStop}%
\bibitem [{\citenamefont {Dekker}\ \emph {et~al.}(2021)\citenamefont {Dekker},
  \citenamefont {Peerbooms}, \citenamefont {Zimmer}, \citenamefont {Ng},\ and\
  \citenamefont {Ando}}]{Dekker:2021bos}%
  \BibitemOpen
  \bibfield  {author} {\bibinfo {author} {\bibfnamefont {A.}~\bibnamefont
  {Dekker}}, \bibinfo {author} {\bibfnamefont {E.}~\bibnamefont {Peerbooms}},
  \bibinfo {author} {\bibfnamefont {F.}~\bibnamefont {Zimmer}}, \bibinfo
  {author} {\bibfnamefont {K.~C.~Y.}\ \bibnamefont {Ng}},\ and\ \bibinfo
  {author} {\bibfnamefont {S.}~\bibnamefont {Ando}},\ }\bibfield  {title}
  {\bibinfo {title} {{Searches for sterile neutrinos and axionlike particles
  from the Galactic halo with eROSITA}},\ }\href
  {https://doi.org/10.1103/PhysRevD.104.023021} {\bibfield  {journal} {\bibinfo
   {journal} {Phys. Rev. D}\ }\textbf {\bibinfo {volume} {104}},\ \bibinfo
  {pages} {023021} (\bibinfo {year} {2021})},\ \Eprint
  {https://arxiv.org/abs/2103.13241} {arXiv:2103.13241 [astro-ph.HE]}
  \BibitemShut {NoStop}%
\bibitem [{\citenamefont {Bulbul}\ \emph {et~al.}(2014)\citenamefont {Bulbul},
  \citenamefont {Markevitch}, \citenamefont {Foster}, \citenamefont {Smith},
  \citenamefont {Loewenstein},\ and\ \citenamefont {Randall}}]{Bulbul:2014sua}%
  \BibitemOpen
  \bibfield  {author} {\bibinfo {author} {\bibfnamefont {E.}~\bibnamefont
  {Bulbul}}, \bibinfo {author} {\bibfnamefont {M.}~\bibnamefont {Markevitch}},
  \bibinfo {author} {\bibfnamefont {A.}~\bibnamefont {Foster}}, \bibinfo
  {author} {\bibfnamefont {R.~K.}\ \bibnamefont {Smith}}, \bibinfo {author}
  {\bibfnamefont {M.}~\bibnamefont {Loewenstein}},\ and\ \bibinfo {author}
  {\bibfnamefont {S.~W.}\ \bibnamefont {Randall}},\ }\bibfield  {title}
  {\bibinfo {title} {{Detection of An Unidentified Emission Line in the Stacked
  X-ray spectrum of Galaxy Clusters}},\ }\href
  {https://doi.org/10.1088/0004-637X/789/1/13} {\bibfield  {journal} {\bibinfo
  {journal} {Astrophys. J.}\ }\textbf {\bibinfo {volume} {789}},\ \bibinfo
  {pages} {13} (\bibinfo {year} {2014})},\ \Eprint
  {https://arxiv.org/abs/1402.2301} {arXiv:1402.2301 [astro-ph.CO]}
  \BibitemShut {NoStop}%
\bibitem [{\citenamefont {Boyarsky}\ \emph {et~al.}(2014)\citenamefont
  {Boyarsky}, \citenamefont {Ruchayskiy}, \citenamefont {Iakubovskyi},\ and\
  \citenamefont {Franse}}]{Boyarsky:2014jta}%
  \BibitemOpen
  \bibfield  {author} {\bibinfo {author} {\bibfnamefont {A.}~\bibnamefont
  {Boyarsky}}, \bibinfo {author} {\bibfnamefont {O.}~\bibnamefont
  {Ruchayskiy}}, \bibinfo {author} {\bibfnamefont {D.}~\bibnamefont
  {Iakubovskyi}},\ and\ \bibinfo {author} {\bibfnamefont {J.}~\bibnamefont
  {Franse}},\ }\bibfield  {title} {\bibinfo {title} {{Unidentified Line in
  X-Ray Spectra of the Andromeda Galaxy and Perseus Galaxy Cluster}},\ }\href
  {https://doi.org/10.1103/PhysRevLett.113.251301} {\bibfield  {journal}
  {\bibinfo  {journal} {Phys. Rev. Lett.}\ }\textbf {\bibinfo {volume} {113}},\
  \bibinfo {pages} {251301} (\bibinfo {year} {2014})},\ \Eprint
  {https://arxiv.org/abs/1402.4119} {arXiv:1402.4119 [astro-ph.CO]}
  \BibitemShut {NoStop}%
\bibitem [{\citenamefont {Urban}\ \emph {et~al.}(2015)\citenamefont {Urban},
  \citenamefont {Werner}, \citenamefont {Allen}, \citenamefont {Simionescu},
  \citenamefont {Kaastra},\ and\ \citenamefont {Strigari}}]{Urban:2014yda}%
  \BibitemOpen
  \bibfield  {author} {\bibinfo {author} {\bibfnamefont {O.}~\bibnamefont
  {Urban}}, \bibinfo {author} {\bibfnamefont {N.}~\bibnamefont {Werner}},
  \bibinfo {author} {\bibfnamefont {S.~W.}\ \bibnamefont {Allen}}, \bibinfo
  {author} {\bibfnamefont {A.}~\bibnamefont {Simionescu}}, \bibinfo {author}
  {\bibfnamefont {J.~S.}\ \bibnamefont {Kaastra}},\ and\ \bibinfo {author}
  {\bibfnamefont {L.~E.}\ \bibnamefont {Strigari}},\ }\bibfield  {title}
  {\bibinfo {title} {{A Suzaku Search for Dark Matter Emission Lines in the
  X-ray Brightest Galaxy Clusters}},\ }\href
  {https://doi.org/10.1093/mnras/stv1142} {\bibfield  {journal} {\bibinfo
  {journal} {Mon. Not. Roy. Astron. Soc.}\ }\textbf {\bibinfo {volume} {451}},\
  \bibinfo {pages} {2447} (\bibinfo {year} {2015})},\ \Eprint
  {https://arxiv.org/abs/1411.0050} {arXiv:1411.0050 [astro-ph.CO]}
  \BibitemShut {NoStop}%
\bibitem [{\citenamefont {Jeltema}\ and\ \citenamefont
  {Profumo}(2015)}]{Jeltema:2014qfa}%
  \BibitemOpen
  \bibfield  {author} {\bibinfo {author} {\bibfnamefont {T.~E.}\ \bibnamefont
  {Jeltema}}\ and\ \bibinfo {author} {\bibfnamefont {S.}~\bibnamefont
  {Profumo}},\ }\bibfield  {title} {\bibinfo {title} {{Discovery of a 3.5 keV
  line in the Galactic Centre and a critical look at the origin of the line
  across astronomical targets}},\ }\href {https://doi.org/10.1093/mnras/stv768}
  {\bibfield  {journal} {\bibinfo  {journal} {Mon. Not. Roy. Astron. Soc.}\
  }\textbf {\bibinfo {volume} {450}},\ \bibinfo {pages} {2143} (\bibinfo {year}
  {2015})},\ \Eprint {https://arxiv.org/abs/1408.1699} {arXiv:1408.1699
  [astro-ph.HE]} \BibitemShut {NoStop}%
\bibitem [{\citenamefont {Cappelluti}\ \emph {et~al.}(2018)\citenamefont
  {Cappelluti}, \citenamefont {Bulbul}, \citenamefont {Foster}, \citenamefont
  {Natarajan}, \citenamefont {Urry}, \citenamefont {Bautz}, \citenamefont
  {Civano}, \citenamefont {Miller},\ and\ \citenamefont
  {Smith}}]{Cappelluti:2017ywp}%
  \BibitemOpen
  \bibfield  {author} {\bibinfo {author} {\bibfnamefont {N.}~\bibnamefont
  {Cappelluti}}, \bibinfo {author} {\bibfnamefont {E.}~\bibnamefont {Bulbul}},
  \bibinfo {author} {\bibfnamefont {A.}~\bibnamefont {Foster}}, \bibinfo
  {author} {\bibfnamefont {P.}~\bibnamefont {Natarajan}}, \bibinfo {author}
  {\bibfnamefont {M.~C.}\ \bibnamefont {Urry}}, \bibinfo {author}
  {\bibfnamefont {M.~W.}\ \bibnamefont {Bautz}}, \bibinfo {author}
  {\bibfnamefont {F.}~\bibnamefont {Civano}}, \bibinfo {author} {\bibfnamefont
  {E.}~\bibnamefont {Miller}},\ and\ \bibinfo {author} {\bibfnamefont {R.~K.}\
  \bibnamefont {Smith}},\ }\bibfield  {title} {\bibinfo {title} {{Searching for
  the 3.5 keV Line in the Deep Fields with Chandra: the 10 Ms observations}},\
  }\href {https://doi.org/10.3847/1538-4357/aaaa68} {\bibfield  {journal}
  {\bibinfo  {journal} {Astrophys. J.}\ }\textbf {\bibinfo {volume} {854}},\
  \bibinfo {pages} {179} (\bibinfo {year} {2018})},\ \Eprint
  {https://arxiv.org/abs/1701.07932} {arXiv:1701.07932 [astro-ph.CO]}
  \BibitemShut {NoStop}%
\bibitem [{\citenamefont {{Nasa High Energy Astrophysics Science Archive
  Research Center (Heasarc)}}(2014)}]{2014ascl.soft08004N}%
  \BibitemOpen
  \bibfield  {author} {\bibinfo {author} {\bibnamefont {{Nasa High Energy
  Astrophysics Science Archive Research Center (Heasarc)}}},\ }\href@noop {}
  {\bibinfo {title} {{HEAsoft: Unified Release of FTOOLS and XANADU}}},\
  \bibinfo {howpublished} {Astrophysics Source Code Library, record
  ascl:1408.004} (\bibinfo {year} {2014}),\ \Eprint
  {https://arxiv.org/abs/1408.004} {ascl:1408.004} \BibitemShut {NoStop}%
\bibitem [{\citenamefont {Dehnen}\ \emph {et~al.}(2006)\citenamefont {Dehnen},
  \citenamefont {McLaughlin},\ and\ \citenamefont {Sachania}}]{Dehnen:2006cm}%
  \BibitemOpen
  \bibfield  {author} {\bibinfo {author} {\bibfnamefont {W.}~\bibnamefont
  {Dehnen}}, \bibinfo {author} {\bibfnamefont {D.}~\bibnamefont {McLaughlin}},\
  and\ \bibinfo {author} {\bibfnamefont {J.}~\bibnamefont {Sachania}},\
  }\bibfield  {title} {\bibinfo {title} {{The velocity dispersion and mass
  profile of the milky way}},\ }\href
  {https://doi.org/10.1111/j.1365-2966.2006.10404.x} {\bibfield  {journal}
  {\bibinfo  {journal} {Mon. Not. Roy. Astron. Soc.}\ }\textbf {\bibinfo
  {volume} {369}},\ \bibinfo {pages} {1688} (\bibinfo {year} {2006})},\ \Eprint
  {https://arxiv.org/abs/astro-ph/0603825} {arXiv:astro-ph/0603825}
  \BibitemShut {NoStop}%
\bibitem [{\citenamefont {Leung}\ \emph {et~al.}(2022)\citenamefont {Leung},
  \citenamefont {Bovy}, \citenamefont {Mackereth}, \citenamefont {Hunt},
  \citenamefont {Lane},\ and\ \citenamefont {Wilson}}]{Leung:2022dno}%
  \BibitemOpen
  \bibfield  {author} {\bibinfo {author} {\bibfnamefont {H.~W.}\ \bibnamefont
  {Leung}}, \bibinfo {author} {\bibfnamefont {J.}~\bibnamefont {Bovy}},
  \bibinfo {author} {\bibfnamefont {J.~T.}\ \bibnamefont {Mackereth}}, \bibinfo
  {author} {\bibfnamefont {J.~A.~S.}\ \bibnamefont {Hunt}}, \bibinfo {author}
  {\bibfnamefont {R.~R.}\ \bibnamefont {Lane}},\ and\ \bibinfo {author}
  {\bibfnamefont {J.~C.}\ \bibnamefont {Wilson}},\ }\bibfield  {title}
  {\bibinfo {title} {{A measurement of the distance to the Galactic centre
  using the kinematics of bar stars}},\ }\href
  {https://doi.org/10.1093/mnras/stac3529} {\bibfield  {journal} {\bibinfo
  {journal} {Mon. Not. Roy. Astron. Soc.}\ }\textbf {\bibinfo {volume} {519}},\
  \bibinfo {pages} {948} (\bibinfo {year} {2022})},\ \Eprint
  {https://arxiv.org/abs/2204.12551} {arXiv:2204.12551 [astro-ph.GA]}
  \BibitemShut {NoStop}%
\bibitem [{\citenamefont {{Eilers}}\ \emph {et~al.}(2019)\citenamefont
  {{Eilers}}, \citenamefont {{Hogg}}, \citenamefont {{Rix}},\ and\
  \citenamefont {{Ness}}}]{2019ApJ...871..120E}%
  \BibitemOpen
  \bibfield  {author} {\bibinfo {author} {\bibfnamefont {A.-C.}\ \bibnamefont
  {{Eilers}}}, \bibinfo {author} {\bibfnamefont {D.~W.}\ \bibnamefont
  {{Hogg}}}, \bibinfo {author} {\bibfnamefont {H.-W.}\ \bibnamefont {{Rix}}},\
  and\ \bibinfo {author} {\bibfnamefont {M.~K.}\ \bibnamefont {{Ness}}},\
  }\bibfield  {title} {\bibinfo {title} {{The Circular Velocity Curve of the
  Milky Way from 5 to 25 kpc}},\ }\href
  {https://doi.org/10.3847/1538-4357/aaf648} {\bibfield  {journal} {\bibinfo
  {journal} {\apj}\ }\textbf {\bibinfo {volume} {871}},\ \bibinfo {eid} {120}
  (\bibinfo {year} {2019})},\ \Eprint {https://arxiv.org/abs/1810.09466}
  {arXiv:1810.09466 [astro-ph.GA]} \BibitemShut {NoStop}%
\bibitem [{\citenamefont {Bonnivard}\ \emph {et~al.}(2015)\citenamefont
  {Bonnivard} \emph {et~al.}}]{Bonnivard:2015xpq}%
  \BibitemOpen
  \bibfield  {author} {\bibinfo {author} {\bibfnamefont {V.}~\bibnamefont
  {Bonnivard}} \emph {et~al.},\ }\bibfield  {title} {\bibinfo {title} {{Dark
  matter annihilation and decay in dwarf spheroidal galaxies: The classical and
  ultrafaint dSphs}},\ }\href {https://doi.org/10.1093/mnras/stv1601}
  {\bibfield  {journal} {\bibinfo  {journal} {Mon. Not. Roy. Astron. Soc.}\
  }\textbf {\bibinfo {volume} {453}},\ \bibinfo {pages} {849} (\bibinfo {year}
  {2015})},\ \Eprint {https://arxiv.org/abs/1504.02048} {arXiv:1504.02048
  [astro-ph.HE]} \BibitemShut {NoStop}%
\bibitem [{\citenamefont {Lisanti}\ \emph
  {et~al.}(2018{\natexlab{a}})\citenamefont {Lisanti}, \citenamefont
  {Mishra-Sharma}, \citenamefont {Rodd},\ and\ \citenamefont
  {Safdi}}]{Lisanti:2017qlb}%
  \BibitemOpen
  \bibfield  {author} {\bibinfo {author} {\bibfnamefont {M.}~\bibnamefont
  {Lisanti}}, \bibinfo {author} {\bibfnamefont {S.}~\bibnamefont
  {Mishra-Sharma}}, \bibinfo {author} {\bibfnamefont {N.~L.}\ \bibnamefont
  {Rodd}},\ and\ \bibinfo {author} {\bibfnamefont {B.~R.}\ \bibnamefont
  {Safdi}},\ }\bibfield  {title} {\bibinfo {title} {{Search for Dark Matter
  Annihilation in Galaxy Groups}},\ }\href
  {https://doi.org/10.1103/PhysRevLett.120.101101} {\bibfield  {journal}
  {\bibinfo  {journal} {Phys. Rev. Lett.}\ }\textbf {\bibinfo {volume} {120}},\
  \bibinfo {pages} {101101} (\bibinfo {year} {2018}{\natexlab{a}})},\ \Eprint
  {https://arxiv.org/abs/1708.09385} {arXiv:1708.09385 [astro-ph.CO]}
  \BibitemShut {NoStop}%
\bibitem [{\citenamefont {Lisanti}\ \emph
  {et~al.}(2018{\natexlab{b}})\citenamefont {Lisanti}, \citenamefont
  {Mishra-Sharma}, \citenamefont {Rodd}, \citenamefont {Safdi},\ and\
  \citenamefont {Wechsler}}]{Lisanti:2017qoz}%
  \BibitemOpen
  \bibfield  {author} {\bibinfo {author} {\bibfnamefont {M.}~\bibnamefont
  {Lisanti}}, \bibinfo {author} {\bibfnamefont {S.}~\bibnamefont
  {Mishra-Sharma}}, \bibinfo {author} {\bibfnamefont {N.~L.}\ \bibnamefont
  {Rodd}}, \bibinfo {author} {\bibfnamefont {B.~R.}\ \bibnamefont {Safdi}},\
  and\ \bibinfo {author} {\bibfnamefont {R.~H.}\ \bibnamefont {Wechsler}},\
  }\bibfield  {title} {\bibinfo {title} {{Mapping Extragalactic Dark Matter
  Annihilation with Galaxy Surveys: A Systematic Study of Stacked Group
  Searches}},\ }\href {https://doi.org/10.1103/PhysRevD.97.063005} {\bibfield
  {journal} {\bibinfo  {journal} {Phys. Rev.}\ }\textbf {\bibinfo {volume}
  {D97}},\ \bibinfo {pages} {063005} (\bibinfo {year} {2018}{\natexlab{b}})},\
  \Eprint {https://arxiv.org/abs/1709.00416} {arXiv:1709.00416 [astro-ph.CO]}
  \BibitemShut {NoStop}%
\end{thebibliography}%

\clearpage

\onecolumngrid
\begin{center}
  \textbf{\large 
  Limits from the grave: resurrecting Hitomi for decaying dark matter and \\
  forecasting leading sensitivity for XRISM
  }\\[.2cm]
  \vspace{0.05in}
  {Christopher Dessert, Orion Ning, Nicholas L. Rodd, and Benjamin R. Safdi}
\end{center}

\twocolumngrid
\setcounter{equation}{0}
\setcounter{figure}{0}
\setcounter{table}{0}
\setcounter{section}{0}
\setcounter{page}{1}
\makeatletter
\renewcommand{\theequation}{S\arabic{equation}}
\renewcommand{\thefigure}{S\arabic{figure}}
\renewcommand{\thetable}{S\arabic{table}}

\onecolumngrid

This Supplementary Material (SM) is organized as follows. In Sec.~\ref{sec:reduction} we outline our methodology for our reduction of the Hitomi blank sky observations.  Section~\ref{sec:Doppler} describes how we account for the Doppler shifting and broadening of our signal, while Sec.~\ref{Sec:SM_fig} presents supplementary figures.

\section{Hitomi Data Reduction}
\label{sec:reduction}

We use HEASOFT version 6.25~\cite{2014ascl.soft08004N} to reduce the Hitomi data sets. We download the observations in Tab.~\ref{tab:obs} from the HEASARC Browse interface. We first screen the electrical cross-talk events from the events file. We extract the spectrum from all pixels except the calibration pixel and grades $\geq 1$ using \texttt{xselect}. We compute the redistribution matrix file (RMF), which models the energy resolution of the instrument, using the task \texttt{sxsmkrmf}. We compute the ancillary response file (ARF), which models the energy-dependent effective area, using the task \texttt{ahexpmap} to create an exposure map and \texttt{aharfgen} to generate the ARF over the defined exposure map. Note that the GV was closed during all the Hitomi observations, which reduced the effective area to near-zero below a couple keV and moderately at higher energies relative to the expectations for science observations.

\begin{table}[hb]
\begin{tabular}{llccc}
ObsID & Target     & $t_{\rm exp}$ [ks] & $l$ [$^\circ$] & $b$ [$^\circ$] \\ \hline \hline
100042010      & IGR J16318-4848 & $73.4$ & $335.49647$ & $-0.34239$           \\
100042020      & IGR J16318-4848 & $68.2$ & $335.49628$ & $-0.34266$           \\
100042030      & IGR J16318-4848 & $40.2$ & $335.49237$ & $-0.33792$           \\
100042040      & IGR J16318-4848 & $68.0$ & $335.55257$ & $-0.39531$           \\
100043010      & RX J1856.5-3754 & $40.8$ & $358.59830$ & $-17.21184$          \\
100043020      & RX J1856.5-3754 & $39.5$ & $358.59849$ & $-17.21274$          \\
100043040      & RX J1856.5-3754 & $47.0$ & $358.59845$ & $-17.21224$          \\
100043050      & RX J1856.5-3754 & $41.0$ & $358.59814$ & $-17.21238$          \\
100043060      & RX J1856.5-3754 & $45.0$ & $358.59823$ & $-17.21243$         
\end{tabular}
\caption{\label{tab:obs} Information on the observations used in the Hitomi analysis. The first column is the observation ID, while the second is the target name. The third column gives the exposure time in [ks]. The fourth and fifth columns show the Galactic coordinates of the observation pointing direction, which can in general be different than the target location. Note that ObsID 100043030 is not included in this analysis because the $^{55}$Fe filter was open, increasing background rates by a factor $\sim$500.}
\end{table}

\section{Accounting for Doppler broadening and Doppler shifting}
\label{sec:Doppler}

As discussed in the main Letter, given the unprecedented energy resolution of Hitomi and XRISM it is crucial to account for the intrinsic line width of the signal, before convolution with the detector response, due to Doppler broadening from the DM velocity dispersion.  Additionally, we must account for the Doppler shift of the signal due to the line-of-sight velocity between the Galactic frame and the solar frame.  Here we provide additional details of how we model these effects.

We may separate the finite-velocity effects into that of the Doppler shift due to the solar velocity and that of the Doppler broadening due to the DM velocity dispersion. We begin with a discussion of the former. Suppose that in the Galactic rest frame the energy distribution function, accounting for the Doppler broadening effects, is given by $f_{\rm gal}(E; m_\chi, \ell,b)$, where $(\ell,b)$ specify the observation direction in Galactic coordinates. We are interested in the boosted energy distribution function $f(E; m_\chi, \ell,b)$ in the solar frame, with the solar frame boosted with respect to the Galactic frame by ${\bf v}_\odot$.  These two distributions are straightforwardly related in the non-relativistic limit by
\begin{equation}
f(E; m_\chi, \ell,b) = f_{\rm gal}\left(E ( 1  - {\bf \hat n} \cdot {\bf v}_\odot /c ) ; m_\chi, \ell,b\right)\!,
\end{equation}
where
\begin{equation}
{\bf \hat n} = (\cos b \cos \ell, \cos b \sin \ell, \sin b)
\end{equation}
is a unit vector that points in the direction of the observation of interest.  Note that if we neglect the Doppler broadening, then $f_{\rm gal}(E; m_\chi, \ell, b) = \delta(E - m_\chi/2)$, which implies that in the solar frame $f(E; m_\chi,\ell,b) = \delta\left(E - m_\chi/2(1 + {\bf \hat n} \cdot {\bf v}_\odot / c ) \right)$.

We now turn to the computation of the Doppler broadening effect due to the finite velocity dispersion of the DM. Under the assumptions of a homogeneous and isotropic DM velocity distribution for a collisionless DM species in gravitational equilibrium with a gravitational potential, the DM velocity distribution at a distance $r$ from the GC is given approximately by (see, {\it e.g.}, Ref.~\cite{Dehnen:2006cm})
\begin{equation}
f({\bf v};r) = {1 \over \pi^{3/2} v_0^3} e^{-{\bf v}^2 / v_0^2},
\end{equation}
where 
\begin{equation}
v_0^2 = {2 \over \gamma - 2 \alpha} V_c^2(r) 
\end{equation}
implicitly depends on the distance $r$ from the GC.
Here, $V_c(r)$ is the circular velocity as a function of the radius $r$, which may be computed by 
\begin{equation}
V_c^2(r) = {G M_{\rm tot}(r) \over r},
\end{equation}
with $M_{\rm tot}(r)$ the mass enclosed within radius $r$.   The coefficients $\alpha$ and $\gamma$ are defined by
\begin{equation}
\alpha \equiv r {\partial_r V_c(r) \over V_c(r)}, \hspace{0.5cm}
\gamma \equiv - r {\partial_r \rho_{\scriptscriptstyle \textrm{DM}}(r) \over \rho_{\scriptscriptstyle \textrm{DM}}(r)},
\end{equation}
with $\rho_{\scriptscriptstyle \textrm{DM}}(r)$ the DM density profile as a function of distance from the GC.

Given the full velocity distribution $f({\bf v}; r)$, we need to compute the distribution of DM velocities projected along the line-of-sight, $v = {\bf v} \cdot {\bf \hat n}$.  Given that the velocity distribution is isotropic, this is simply given by
\begin{equation}
f(v; r) = {1 \over \sqrt{\pi} v_0} e^{-v^2 / v_0^2}.
\end{equation}

As discussed in the main body, we model the DM density as an NFW profile
\begin{equation}
\rho_{\scriptscriptstyle \textrm{DM}}(r) = {\rho_s \over r/r_s (1 + r/r_s)^2},
\end{equation}
and for our fiducial analysis we take, following Ref.~\cite{Foster:2021ngm}, $\rho_s = 6.6 \times 10^6$ $M_\odot / {\rm kpc}^3$ and $r_s = 19.1$ kpc (see Ref.~\cite{Safdi:2022xkm} for an extended discussion of the Milky Way DM density profile). (Note that with this choice the local DM density at the solar radius is $\sim$0.29 GeV/cm$^3$, where the distance of the Sun from the GC is $r_\odot \simeq 8.23$ kpc~\cite{Leung:2022dno}.)  To calculate $V_c(r)$ and $\gamma$ we use the best-fit circular velocity model from Ref.~\cite{2019ApJ...871..120E}, which fit a Galactic potential model consisting of disk, bulge, and halo components to rotation curve data inferred for a sample of red giant stars with 6D phase-space measurements.  Note that at the solar location they find $V_c \simeq 230$ km/s and $\alpha \simeq 0$, meaning that the rotation curve is roughly flat.  On the other hand, $\gamma$ varies between roughly $1$ in the inner Galaxy to $3$ in the outer Galaxy.

To compute the energy distribution function $f(E; m_\chi,\ell,b)$ for two-body decays we must compute a weighted integral accounting for the photons produced along the line of sight in the direction ${\bf \hat n}$ over distance $s$ away from the Sun.  Note that at a given distance $s$ from the Sun in the direction ${\bf \hat n}$ the distance squared to the GC is
\begin{equation}
r^2(s; \ell,b) = r_\odot^2 + s^2 - 2 \cos b \cos \ell \,r_\odot s.
\end{equation}
With that in mind, the Galactic frame energy distribution function may be computed by
\begin{equation}
f(E; m_\chi,\ell,b) = {4 c \over m_\chi} {\int_0^{\infty} ds \, \rho_{\scriptscriptstyle \textrm{DM}}(r) f\big(v(E); r\big) \over \int_0^{\infty}  ds \, \rho_{\scriptscriptstyle \textrm{DM}}(r)},
\label{eq:f_E}
\end{equation}
where the dependence of $r$ on $s$, $\ell$, and $b$ is implicit above and where $|v(E)| / c = {2 \over m_\chi} E - 1$.  Note that in practice the integrals above are cut-off at distances of order the virial radius.  Also note that the extra factor of two in Eq.~\eqref{eq:f_E} is needed to account for both positive and negative $v$.  For simplicity, in this work we assume $v_0 = 220$ km/s everywhere in the Galaxy; we have verified that more realistic $v_0(r)$ profiles yield nearly identical results after convolution with the detector response.

\section{Supplementary Figures}
\label{Sec:SM_fig}

\begin{figure}[!htb]
\centering
\includegraphics[width=0.5\textwidth]{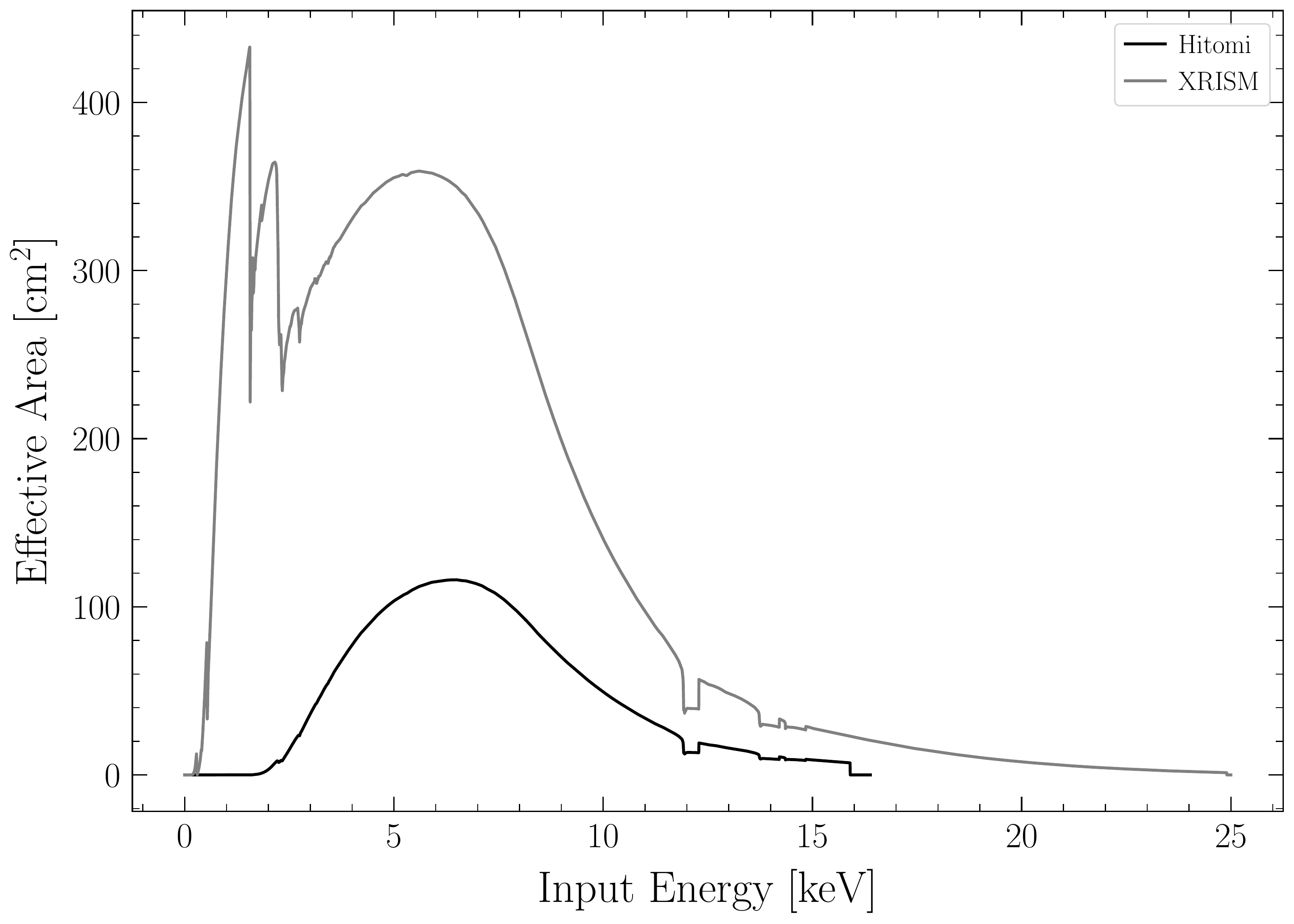}
\vspace{-0.4cm}
\caption{An illustration of the effective area of the Hitomi Soft X-Ray Spectrometer (SXS) as computed from observation 100043010, as well as XRISM {\it Resolve}'s instrument. In our analysis we restrict to input energies between 1 and 15.1 keV. Note the large difference between Hitomi and XRISM is driven by the gate valve being open during Hitomi's operation.}
\label{fig:Aeff}
\vspace{-0.4cm}
\end{figure}

\begin{figure}[!htb]
\centering
\includegraphics[width=0.5\textwidth]{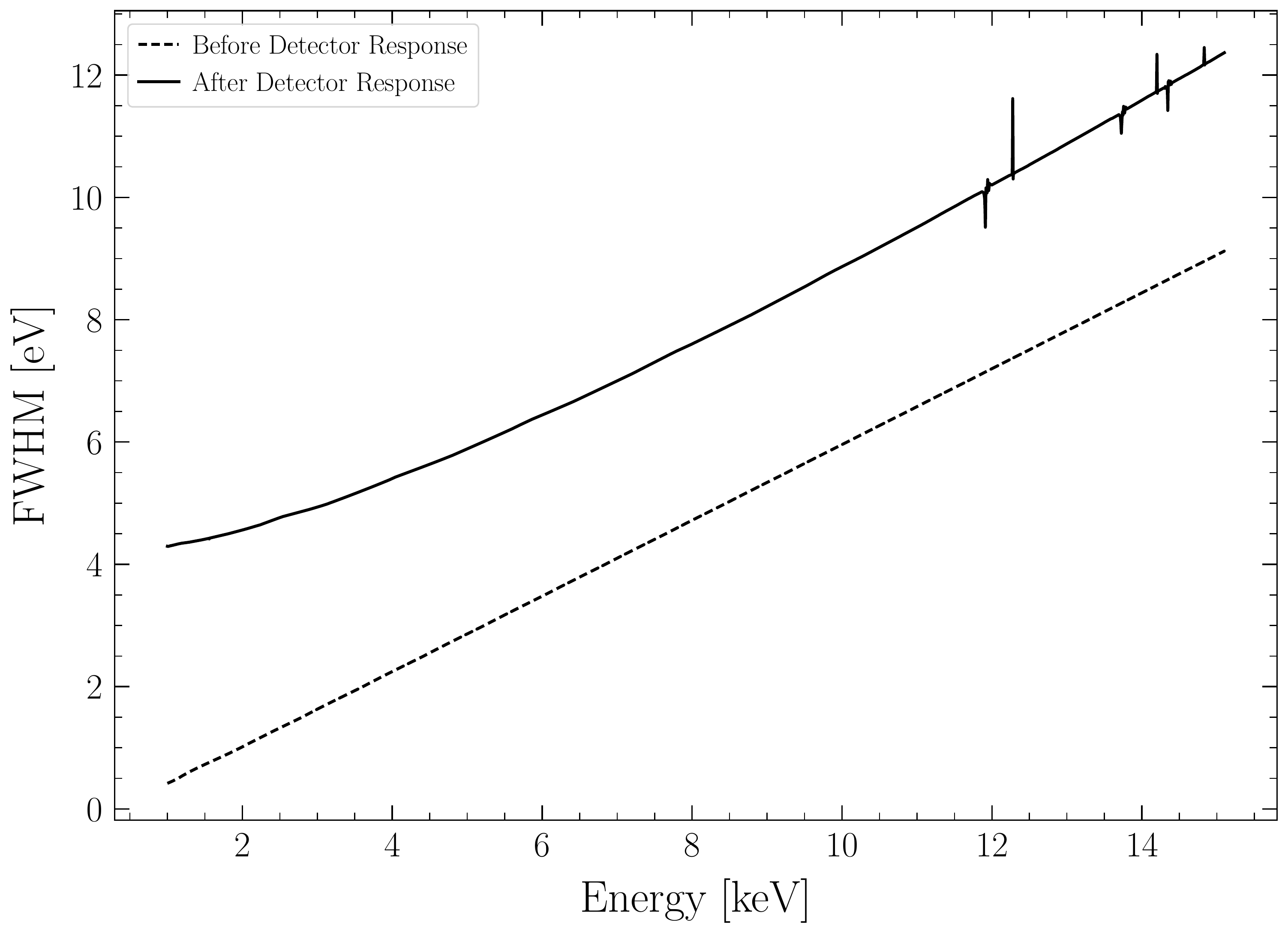}
\vspace{-0.4cm}
\caption{The FWHM of the DM decay signal as a function of energy, computed using the stacked RX J1856.5-3754 data.  The curve before the detector response illustrates the width of the intrinsic signal from Doppler broadening in the galaxy, while the curve after the detector response is at higher values because of the SXS energy resolution.  The jagged regions near and above 12 keV are from discontinuities in the instrument response at those energies.  }
\label{fig:FWHM}
\vspace{-0.4cm}
\end{figure}

\begin{figure}[!t]
\centering
\includegraphics[width=0.5\textwidth]{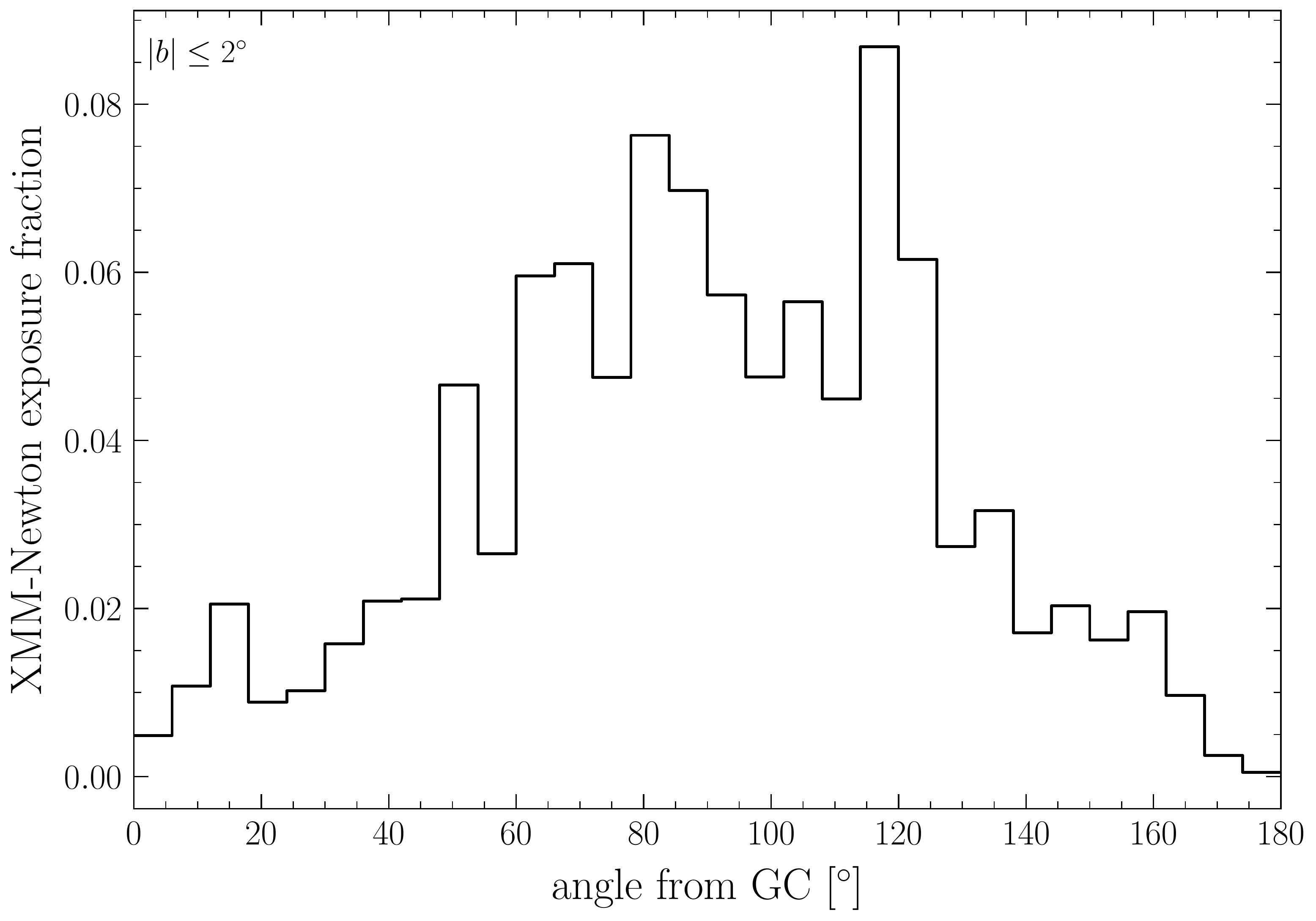}
\vspace{-0.4cm}
\caption{The distribution of exposure time per XRISM-analysis annuli (note that $|b| \leq 2^\circ$ is masked) from the ensemble of all of the XMM-Newton observations passing the quality cuts in Ref.~\cite{Foster:2021ngm}.  We use this exposure distribution when projecting the future distribution of exposures from XRISM.}
\label{fig:xmm_exposure}
\vspace{-0.4cm}
\end{figure}

\begin{figure}[!t]
\centering
\includegraphics[width=0.5\textwidth]{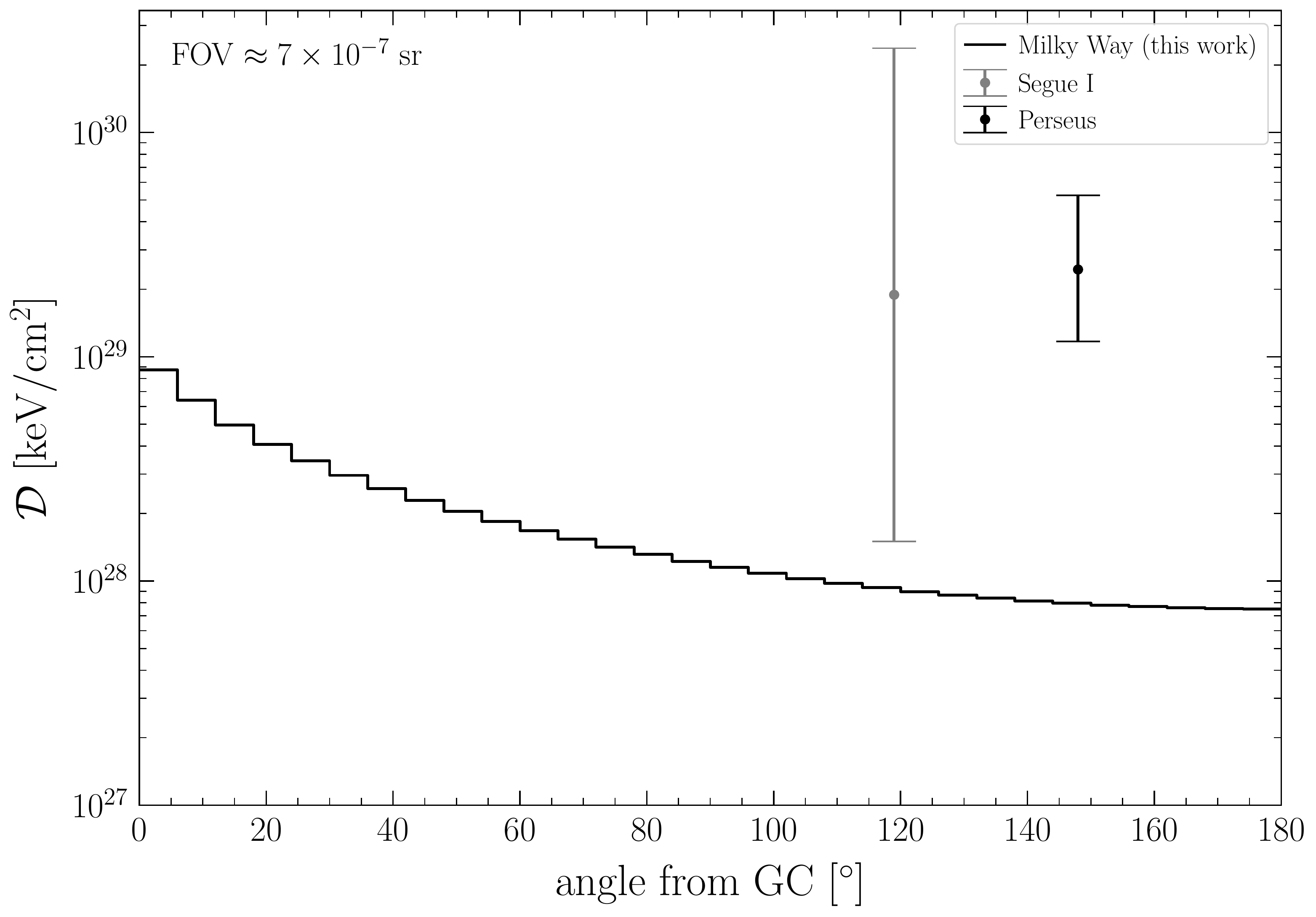}
\vspace{-0.4cm}
\caption{The ${\cal D}$-factor profile for the Milky Way using our canonical NFW DM density profile for the Galaxy in our XRISM-analysis annuli. We compare these ${\cal D}$-factors to those estimated from the Segue I dwarf galaxy and the Perseus cluster, averaged over the FOV of XRISM. We adopt the uncertainties from Ref.~\cite{Bonnivard:2015xpq} for Segue I and compute $1\sigma$ uncertainties on Perseus by taking the mass and concentration parameters provided in Refs.~\cite{Lisanti:2017qlb,Lisanti:2017qoz} to be at their upper and lower 1$\sigma$ values. }
\label{fig:D}
\vspace{-0.4cm}
\end{figure}

\begin{figure}[!t]
\centering
\includegraphics[width=0.5\textwidth]{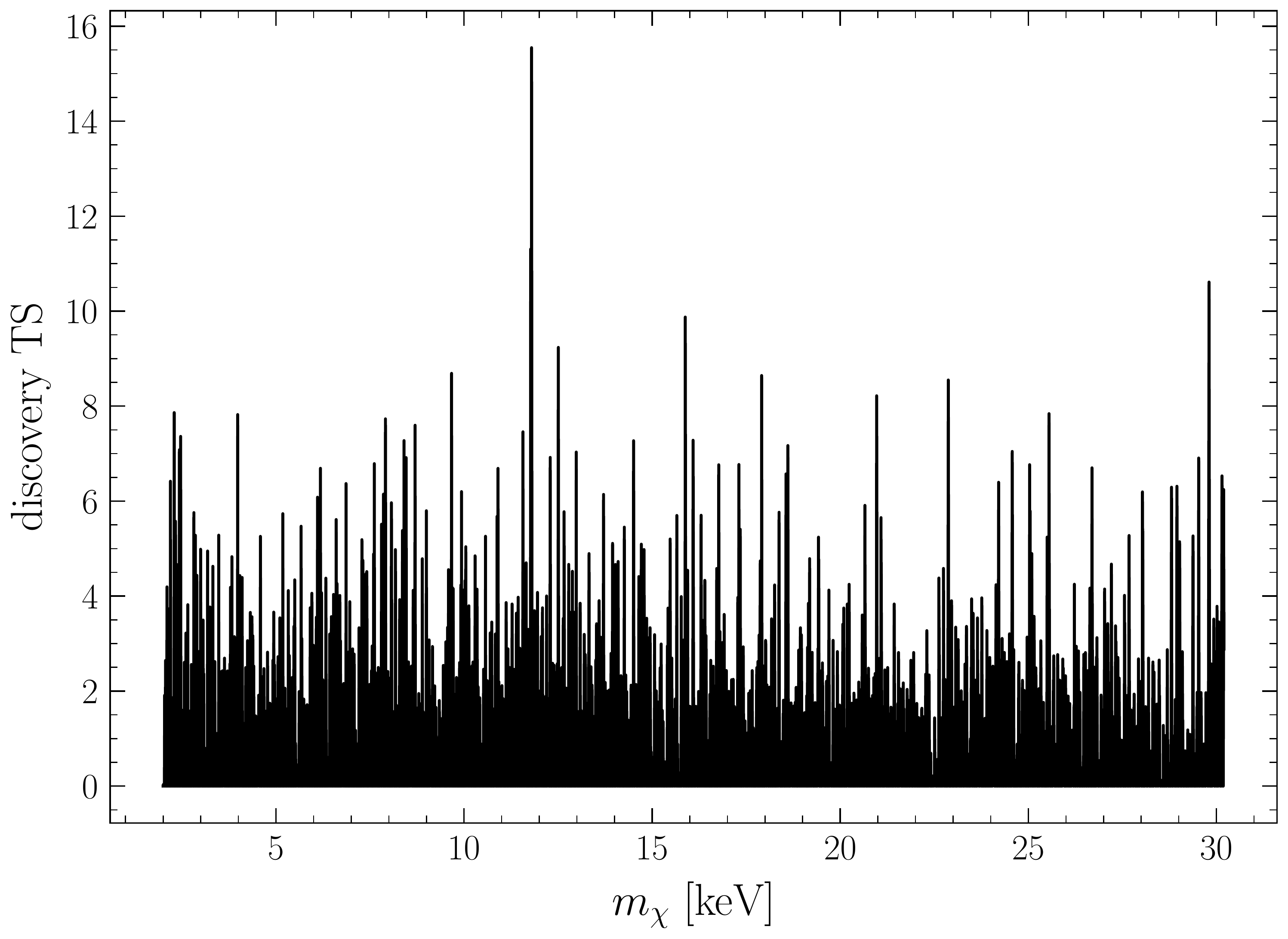}
\vspace{-0.4cm}
\caption{
The discovery TS in favor of the signal model with $\tau_\chi > 0$ as a function of the DM mass $m_\chi$ for our analysis. The highest TS point is at $m_\chi = 11.794$ keV and likely corresponds to the Mn K$\alpha$ line at $\sim$5.8988 keV. 
}
\label{fig:TS_mass}
\vspace{-0.4cm}
\end{figure}

\begin{figure}[!t]
\centering
\includegraphics[width=0.5\textwidth]{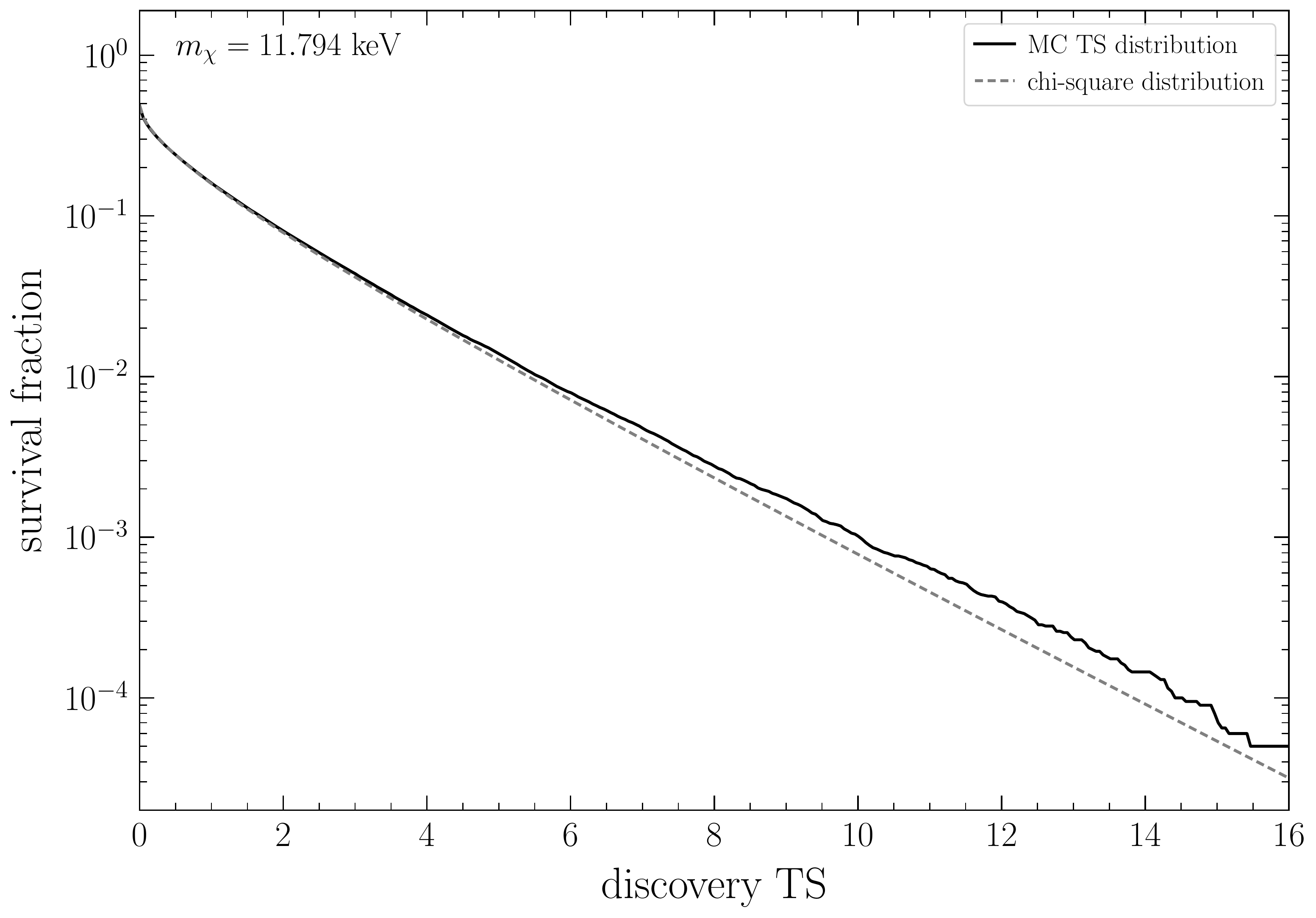}
\vspace{-0.4cm}
\caption{
For our Hitomi mass point that has the largest discovery TS -- $m_\chi = 11.794$ keV -- we simulate a large ensemble ($2 \cdot 10^5$) of null-hypothesis data sets. We analyze each data set for evidence of the signal model and compute the survival fraction of TSs, shown here as in Fig.~\ref{fig:survival}, over the ensemble. We compare this distribution to one-half the survival fraction for the chi-square distribution with one degree of freedom. The two distributions agree out to at least ${\rm TS} \simeq 16$, which justifies our assumption in the Hitomi analyses that the TSs are chi-square distributed, since for the other mass points the TS excursions are smaller than 16.
}
\label{fig:MC}
\vspace{-0.4cm}
\end{figure}

\begin{figure}[!t]
\centering
\includegraphics[width=0.5\textwidth]{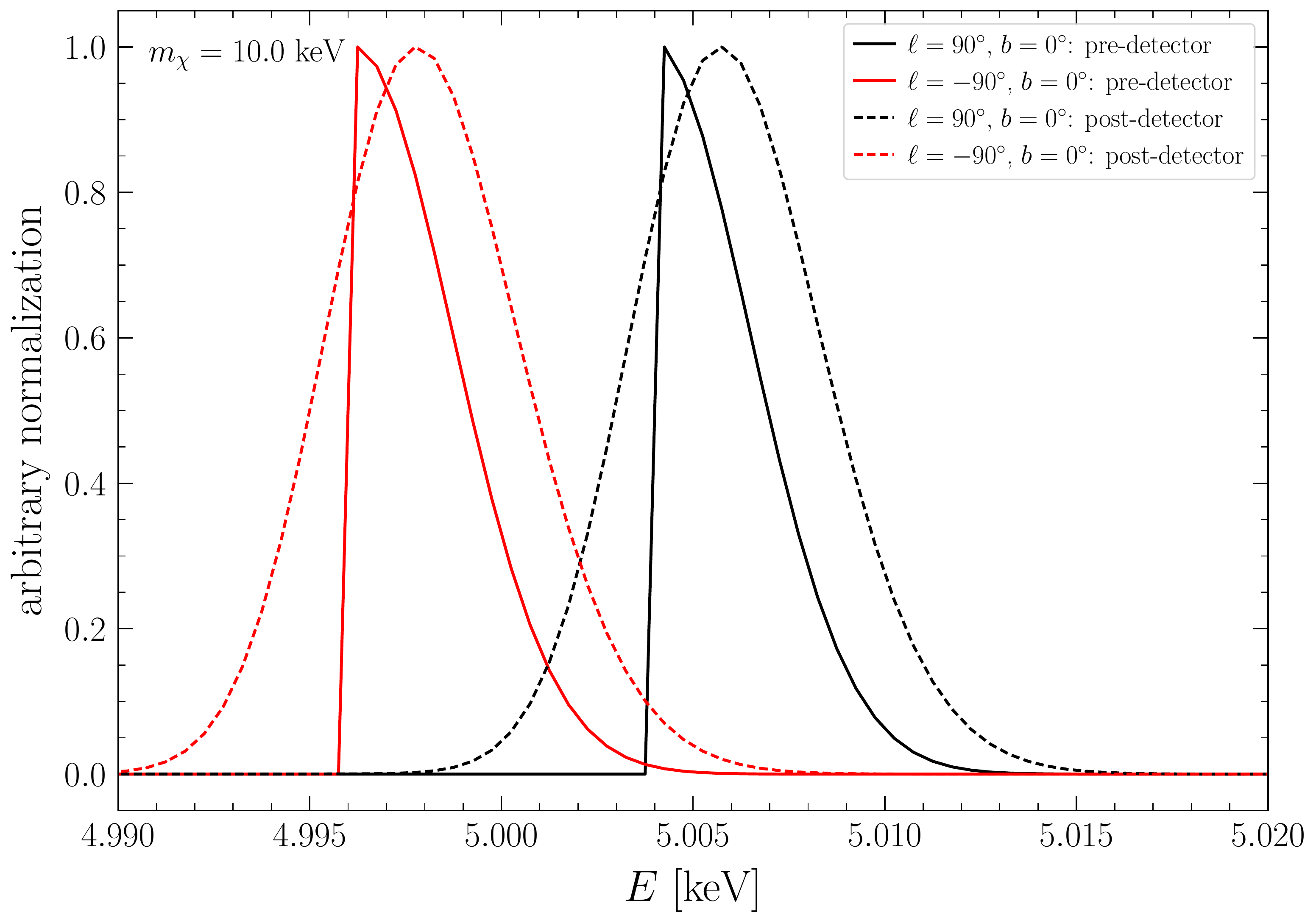}
\vspace{-0.4cm}
\caption{
An illustration of the Doppler broadening and Doppler shifting of the DM decay signal for an $m_\chi = 10.0$ keV decaying DM candidate for observations pointing in the indicated directions.  We show the signal shapes before and after convolving with the instrument response. Note that the signal at positive $\ell$ is shifted to higher energies since the Sun is traveling in the $+{\bf \hat y}$ direction due to the rotation of the disk, while that at negative $\ell$ is at lower energies.  Given the unprecedented energy resolution of XRISM it will be important to account for the Doppler effects self consistently in analyses of their future BSO data for DM decay.  
}
\label{fig:Doppler}
\vspace{-0.4cm}
\end{figure}

\begin{figure}[!t]
\centering
\includegraphics[width=0.45\textwidth]{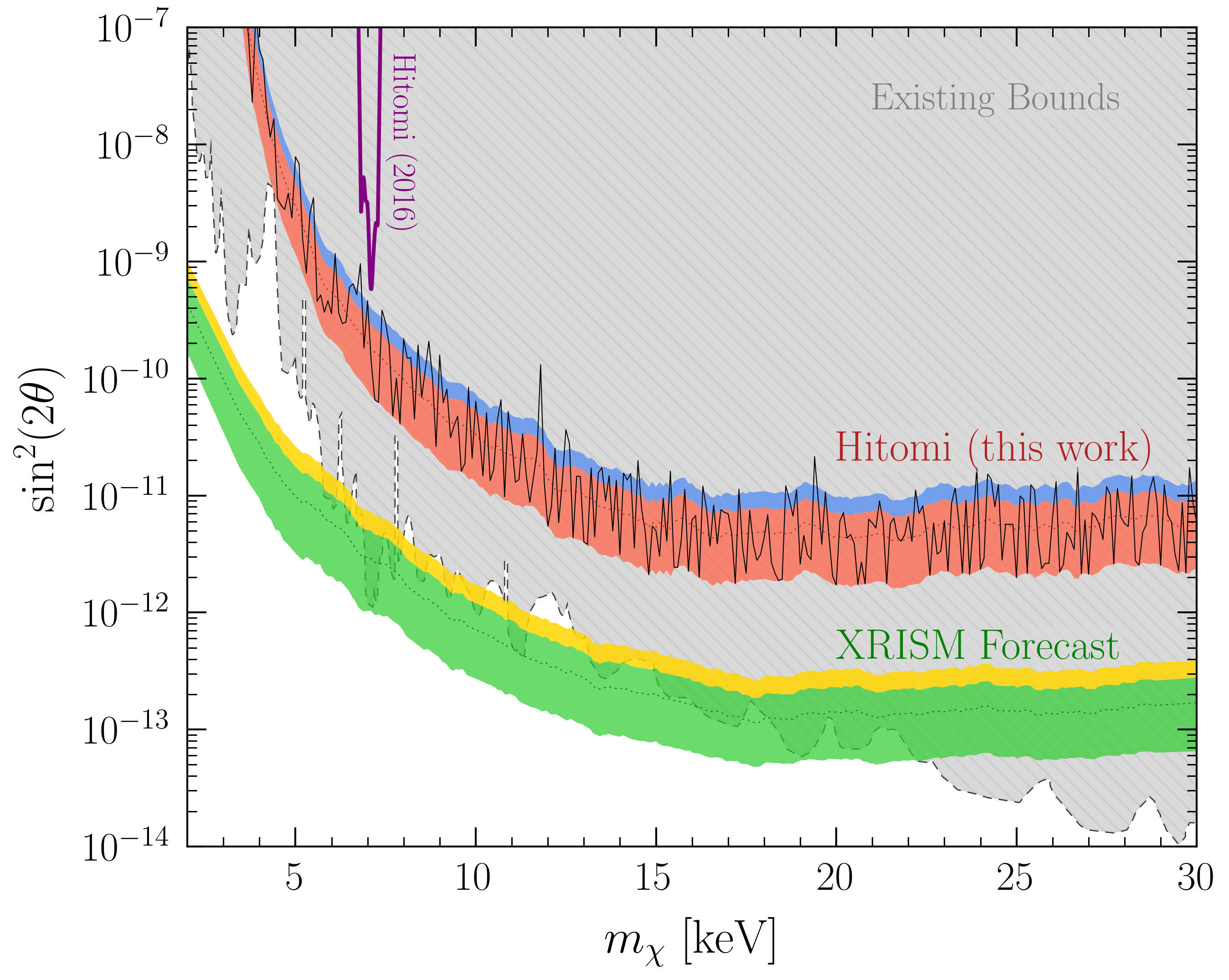} 
\hspace{0.5cm}
\includegraphics[width=0.45\textwidth]{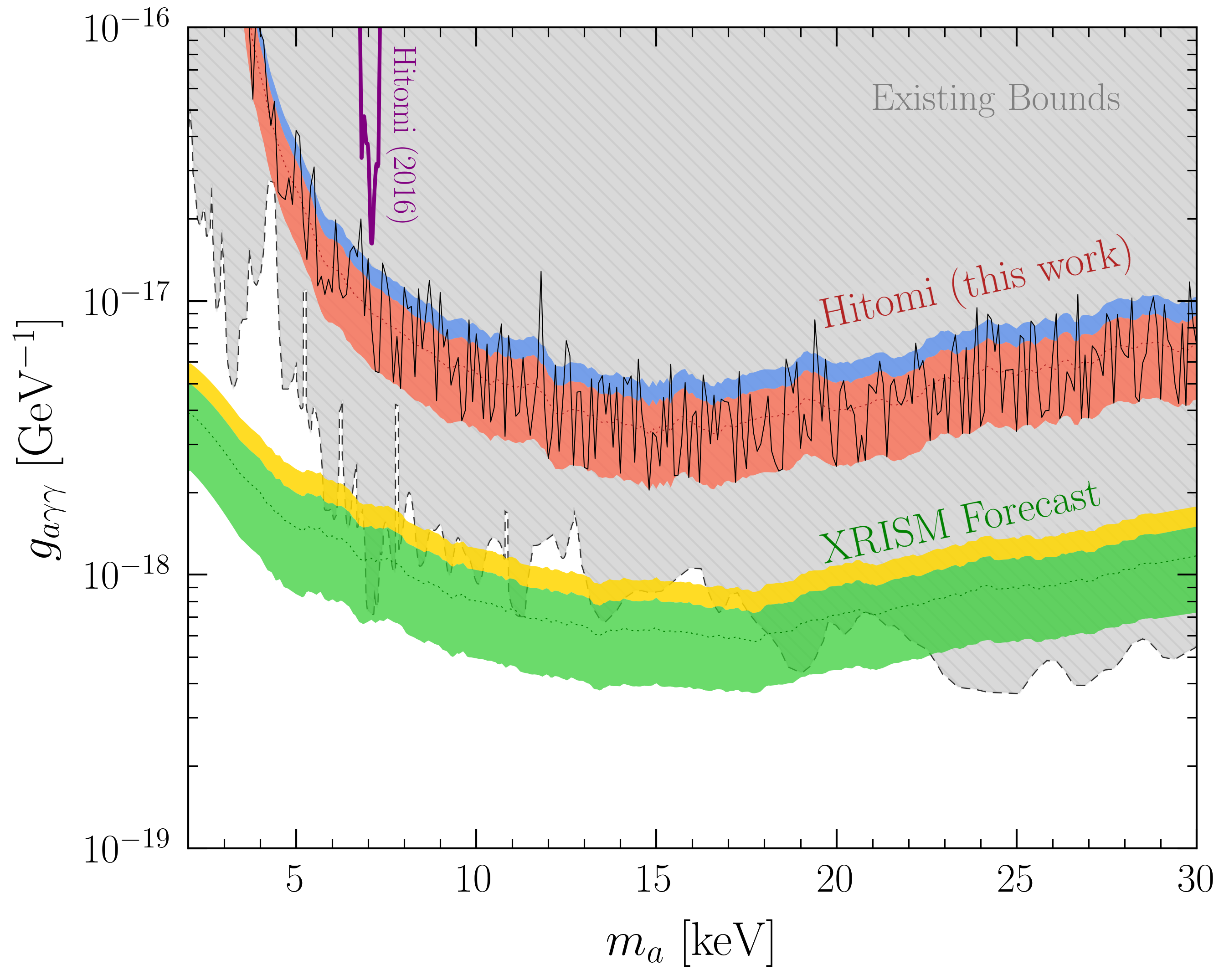}
\vspace{-0.4cm}
\caption{
The results in Fig.~\ref{fig:Limit} reinterpreted in the context of specific particle physics models for DM. In the left panel we show the constraints on the sterile neutrino DM parameter space for sterile neutrinos of mass $m_\chi$ with sterile-active mixing parameter $\sin^2(2\theta)$, while in the right panel we illustrate the axion parameter space where 
the decay of axion DM $a \to \gamma \gamma$ is controlled by the axion-photon coupling $g_{a\gamma\gamma}$.
}
\label{fig:Lifetime+Axion}
\vspace{-0.4cm}
\end{figure}

Here we include supplementary figures that provide further context for the conclusions in the main Letter, including:
\begin{itemize}
\item Fig.~\ref{fig:Aeff}: The Hitomi and XRISM effective areas from our analysis.
\item Fig.~\ref{fig:FWHM}: The FWHM of the putative signal before and after the Hitomi instrumental response.
\item Fig.~\ref{fig:xmm_exposure}: The XMM-Newton exposure distribition, which we use in projecting XRISM sensitivity. 
\item Fig.~\ref{fig:D}: The ${\mathcal D}$-factor profile in the Milky Way and, as examples, the Segue I dwarf galaxy and the Perseus cluster.
\item Fig.~\ref{fig:TS_mass}: The discovery TSs as a function of the DM mass.
\item Fig.~\ref{fig:MC}: The distribution of expected TSs under the null from MC relative to the chi-square expectation.
\item Fig.~\ref{fig:Doppler}: The signal templates accounting for Doppler shifting and broadening towards the two targets for the Hitomi analysis.
\item Fig.~\ref{fig:Lifetime+Axion}: Our limits reinterpreted in terms of sterile neutrino and ALP DM.
\end{itemize}

\end{document}